\documentclass[12pt,preprint]{aastex}

\shorttitle{Monitoring S5 1803+78}
\shortauthors{Nesci et al.}

\input epsf.sty
\begin{document}


\title{Optical and Radio monitoring of S5 1803+784}


\author{R. Nesci, E. Massaro, M. Maesano, F. Montagni, S. Sclavi}
\affil{Dipartimento di Fisica, Universit\'a La Sapienza,
P.le A. Moro 2, 00185, Roma, ITALY}
\email{Roberto.Nesci@uniroma1.it}

\author{T. Venturi, D. Dallacasa}
\affil{Istituto di Radioastronomia CNR
   via P. Gobetti 101, I-40129 Bologna, Italy}
\email{tventuri@ira.bo.cnr.it}

\and

\author{F. D'Alessio}
\affil{Osservatorio Astronomico di Roma, via di Frascati 33, Monteporzio
    Catone I-00040, Italy}
\email{dalessio@coma.mporzio.astro.it}




\begin{abstract}

The optical (BVRI) and radio (8.4 GHz) light curves of S5 1803+784 on a time
span of nearly 6 years are presented and discussed.
The optical light curve showed an overall variation greater than 3 mag, and the
largest changes occured in three strong flares.
No periodicity was found in the light curve on time scales up to
a year. The variability in the
radio band is very different, and shows moderate oscillations around
an average constant flux density rather than relevant flares, with a
maximum amplitude of $\sim$30\%, without a simultaneous correspondence
between optical and radio luminosity.
The optical spectral energy distribution was always well fitted
by a power law. The spectral index shows small variations and there is indication
of a positive correlation with the source luminosity.
Possible explanations of the
source behaviour are discussed in the framework of current models.

\end{abstract}




\section{Introduction}
The radio source S5 1803+784 was classified as a BL Lac object by Biermann
et al. (1981). A redshift estimate (0.680) was derived from a weak MgII line by
Lawrence et al. (1996) and recently confirmed by Rector \& Stocke (2001).
From the available flux measurements, although non simultaneous, the overall
energy distribution from radio to the X rays (in the Log ($\nu F_{\nu}$) $vs$
Log ($\nu$) plot) peaks in the IRAS band  (Giommi et al. 1995), so that the 
source is classified  as a Low Energy Peaked BL Lac (LBL) object
in the scheme proposed by Padovani \& Giommi (1995).
Like other sources of this class, one would expect in the optical band a
remarkable variability with a correlation between the spectral slope and
the luminosity (see e.g. Massaro et al. 1999; Nesci et al. 1998).
The published photometric data, however, are very few:
Wagner et al. (1990) and Heidt \& Wagner (1996) reported only uncalibrated
luminosity variations relative to a few weeks,
and practically nothing is known about the long term time behaviour of the
source.
Only in the radio band S5 1803+784 was extensively
monitored, both from the group of the University of Michigan (Aller et al.
1998) and from the Metsahovi observatory (Teraesranta et al. 1998),
while several VLBI images have been obtained at different epochs since 1979
(Biermann et al. 1981),
because it is normally used as a reference radio source for geodynamics
studies (see Britzen et al. 1999; Gabuzda 1999, 2000).

In April 1996 we started an optical monitoring program to define the long term
light curve and color variations of this source.
A parallel radio monitoring program at 5 GHz and 8.4 GHz was carried
out starting from January 1996 with the two 32-m antennas located
in Medicina (Bologna, Italy) and Noto (Siracusa, Italy), as part of a
larger project to provide radio lightcurves of several bright BL Lac objects,
possible targets of X-ray observations (Venturi et al. 2001).
In this paper we present the observational results of this monitoring and
discuss some
implications of our measures on the nature of the luminosity changes.
The observational data are described in Section 2; the light-curve
temporal analysis is discussed in Section 3; the optical spectral distribution
is discussed in Section 4; the general discussion of the results is made in
Section 5.

\section{Observations}
\subsection{Optical observations}
The large majority of our observations were made with a 50 cm f/4.5
telescope at the Vallinfreda astronomical station near Rome
(850 m a.s.l.), equipped with a CCD camera based on the Texas TC241 chip.
A minor number of data were collected with a 32 cm f/4.5
telescope located at Greve (Tuscany, 650 m a.s.l.) and a 70 cm f/8.3
telescope at Monte Porzio (Roma, 380 m a.s.l.) both equipped with the same
model of CCD camera, based on the SITe SIA501A back-illuminated chip.
The filter sets of all these telescopes are identical and match the
standard B,V (Johnson) and R,I (Cousins) bandpasses. 
The typical exposure times were 300 s for the V,R,I bands
and 600 s for the B band. When the source was faint
(for R$>$16 mag) exposure times were doubled.

Systematic differences between the data sets of the three observatories
were checked during several observational campaigns of BL Lac objects
using simultaneous, or nearly simultaneous observations, and found to
be negligible.
For S5 1803+784, in particular, we have two nights in common between
Vallinfreda and MontePorzio, two nights between Vallinfreda and Greve,
and one night between Greve and Vallinfreda.
Bias, dark and flat field corrections of the frames were performed with
IRAF tasks, and aperture photometry was made with IRAF-apphot, using an
aperture of 5 arcsec radius. Local background level was estimated from an
annular region concentric to each star.

Magnitudes of S5 1803+784 were estimated differentially with respect
to five nearby reference stars.
The final results, listed in Table 3, are the average of the magnitude differences
with respect with the reference stars, and the quoted error is the standard deviation
of the single measures.
The intercalibration of these five stars, in each filter, was made using the
best 35 nights. Their magnitude differences agreed with a standard deviation
of 0.01 magnitudes in each filter. No evidence of variability of these stars was
detected in all our database.

The zero point calibration for each photometric band
was made linking the brightest reference star (Star A in
Table 1) to the photometric sequences of five other BL Lac objects, namely
PKS 0422+004 (Miller et al. 1983); AO 0235+164 (Smith et al. 1985, McGimsey
et al. 1976, Rieke et al. 1976; 3C66A (Fiorucci \& Tosti 1996); S5 0716+714
(Ghisellini et al. 1997); BL Lacertae (Bertaud et al. 1969, Fiorucci \& Tosti
1996).
For this purpose the best photometric nights were used, with air masses
ranging from 1.0 to 1.5. The number of nights used for each filter is listed (in
round bracketts) in Table 1.

For each filter and night a linear fit of the extinction law  was
derived and used to convert the instrumental magnitudes into the apparent ones.
The final result is the average over the different nights, and the error
is the standard deviation of the mean.
Given the large errors found in deriving the calibration of the reference stars
in the B band, a new calibration of star A  
was made using four Bright Stars (HR 6598, 6636, 6637
and 6811). Their B magnitudes were taken from the Bright Star Catalogue, 5th
Revised Edition, as given at the CDS (Hoffleit \& Warren 1991).
The results are reported in Table 1: the errors quoted for star A
include the zero point uncertainty, while those for the other stars are
the intercalibration standard errors
of the mean, derived from the average of 35 images for each colour.
The finding chart is shown in Fig. 1.

From the digitized Palomar sky survey plates and using our photometric sequence,
we also derived a few approximate historical magnitudes for S5 1803+784, which
are collected in Table 2.

\subsection { Optical light curve }

In total we collected data for 207 days, 85 with full BVRI photometry and
further 49 with VRI photometry, since April 1996 until January 2002.
The whole data set is given in Table 3 and the
R light curve is shown in Fig. 2: typical errors are 0.02-0.03 mag when
the source was bright and 0.06 when very faint.

Until JD 2451350 (Sep. 1999), the source behavior can be simply described
as a monotonic increase of the luminosity, with small (0.3 mag) amplitude 
and relatively fast variations above or below the average level, with three
strong bursts at JD 711, 1127 and 1344 (hereafter we report the JD dates
minus 2450000) of comparable amplitude and time scale.
In the last event the source reached its
highest recorded level (R=14.0). Then a decay phase
started, with still some oscillations overimposed, bringing the flux of
S5 1803+784 at the lowest recorded level (R$\sim$17.4, July 2001). After that
a new rising phase started.
The overall magnitude interval is therefore about 3.4 mag, confirming a
large optical variability, typical of this class of BL Lac objects.

The three largest bursts indicated above are reported enlarged in
Fig. 3, 4 and 5. For ease of comparison, the same scales were used for all three Figs.
The V$-$I colour index is also plotted, with the scale
marked on the right hand side.

The first episode (Fig. 3) was observed only in the decreasing branch and
showed a monotonic decline of 1.4 mag in 20 days, with a mean rate of 0.07
mag/day. Because of the sampling only a lower limit
of 0.036 mag/day can be set to the rate of the rising branch.
In the second one (Fig. 4), one can see a rise by 1.5 mag in 21 days (0.071 mag/day).
The subsequent decrease was initially fast, with approximately the same
time scale of the increase; after, our data indicate a change of the slope.
In the third event (Fig. 5) the source reached its brightest state and
likely it remained so bright for about one month, then a decline phase
started of about 1.7 mag in 78 days.

The greatest variation rate, derived from this inspection of the main
events, is therefore 0.07 mag/day, observed both in rising and in
decaying segments. Given the uneven sampling of the light curve we cannot
exclude rates greater than this estimate, but such rates might have lasted only 
on very short (day) time scale.
We think however that very fast variations, if present, do not represent
the normal behavior of the source and the above estimate can be used to
give constraints to theoretical models of variability mechanisms.

\subsection{ Near Infrared observations}

Observations in the Johnson JHK bands were performed in August 2001 with the
AZT24 telescope at Campo Imperatore (110 cm f/7.9, Cassegrain)
and a PICNIC type HgCdTe 256x256 camera, with a scale of 1 arcsec/pixel. 
Each image was the sum of 6 individual frames of 30s exposure, to avoid saturation.
Preliminary data reduction was made with the PREPROCESS task developed at the
Roma Observatory, and photometry was made with IRAF/daophot using a 4 arcsec
radius aperture. Primary photometric standard stars 
(AS02, AS06, AS32, AS37-1) were taken from Hunt et al. (1998).
All the observations were made at airmass smaller than 1.2; the standard error
of our JHK observations is $\sim$0.05 mag. The results are collected in 
Table 4.
Simultaneous optical photometry was obtained with the Vallinfreda telescope.

\subsection{ Radio observations }

The 4-year radio monitoring of S5 1803+784 was carried out at 5 GHz
and 8.4 GHz with the 32-m antennas located in Medicina (Bologna, Italy) and
Noto (Siracusa, Italy) on monthly basis, starting from January 1996.
The observations were performed by means of ON-OFF measurements
on the target source.
The systematic calibration uncertainty of the 
absolute radio flux is of the order of 4\%,
while the typical inner error of the measurements is 10 - 20 mJy.
Details on the observations and data reduction can
be found in Venturi et al. (2001).
The 8.4 GHz data (the more complete dataset in our radio monitoring) are
shown in Fig. 6, upper panel; for ease of comparison the optical data
are shown in the lower panel of the same figure.

The radio light curve shows a significant variability around 
an average flux density of 3 Jy.
The minimum measured flux was 2.46 $\pm$0.02 Jy on JD 657 and the
maximum was 3.46 $\pm$0.01 on JD 1376.
The sampling is much coarser than the optical one, so that a cross correlation
analysis would not be very meaningful:
a simple eye-comparison of the two light curves
does not suggest anyway a simultaneus correspondence between them.
We note however that
simultaneous radio-optical measurements on July 12 2001 (JD 2103),
when the source was near its minimum optical brightness, gave 
radio flux densities of  F$_{8.4}$ = 2.53 Jy (and F$_{5.0}$ = 2.50 Jy at 5 GHz
on July 17), one of the lowest recorded radio levels in our monitoring. 

The amplitude of the radio variability seems to increase 
from 1997 to 2000: the flux density difference
between maxima and minima going from $\sim$ 0.2 Jy (i.e. $\sim$ 7\%) at
the beginning of 1997 to 0.8 Jy (i.e. $\sim 28$\%) in 1999-2000.

\section{Time scale analysis}

The search for possible time scales in the optical light curve was
performed by means of the Discrete Fourier Transform (DFT) for
unevenly spaced data (Deeming 1975). We also used the
Structure Function (SF) analysis (Simonetti et al. 1985) and the
Jurkevich (1971) method.

The DFT power spectrum is plotted in Fig. 7: the upper panel shows the
light curve spectrum, and the lower panel that of the sampling window.
The former spectrum is clearly dominated by a very prominent maximum
at the lowest frequency due to the long term trend; a few other peaks
are also present, two of them corresponding to the periods of about
218 and 105 days. The first period is nearly the distance between the second
and the third flare. The latter could be just the first harmonic of 218,
given the achievable resolution step. Note that in
the window power spectrum a feature is present at 105 days, and therefore this
frequency in the source power spectrum could be enhanced by the convolution effect.
None of these features in the power spectrum, however, is so prominent
to suggest a stable periodicity.
The window power spectrum shows a marked periodicity at $\sim$29 days, corresponding
to the moon phase period.

Application of the Jurkevich test for the search of periodicities was made
starting with a trial period of 5 days, and increasing it with a 5 days
step, binning into 29 intervals the phase-reduced light curve.
Given our average sampling of about 1 point every 10 days,
shorter time scales cannot be explored with our database, save that
in a few short intervals of denser sampling.
For a better detection of possible time scales, we subtracted from the R
light curve the average trend, derived with the running mean over 5
consecutive values.
The result of the Jurkevich test is shown in Fig. 8, where the
{\it f} parameter is plotted against the trial period.
We recall that {\it f} is defined as 1-V$_B$/V$_T$, where V$_B$ is the sum of the 
variances
of the phase-reduced binned data and V$_T$ is the variance of the whole sample. 
Minima in this plot may be considered as indicators of possible time scales:
a value of {\it f} smaller than $\sim$0.5 is often quoted in the
literature as indicator of a 'true' time scale (e.g. Kidger, Takalo \&
Sillanp\"a\"a 1992). Several minima appear in this plot, the shortest one
around 620 days (close to the lag between the first and the third flare),
but all of them are far from the significancy threshold.
Changing  the number of bins and the step of the trial periods does not change
substantially the positions of the minima, nor their relative depth.

Finally the SF plot, averaged over a logaritmic time window of 0.1,
is shown in Fig. 9. It has been calculated on the flux density values in mJy
and no normalization has been applied.
The SF has an average slope of 0.62$\pm$0.04 in the time lag range 
between 10 and 1000
days. A plateau region is marginally apparent between 20 and 80 days.
No clear periodicity, which would reveal as a local minimum, is present.
On short time scales (days) the SF analysis for S5 1803+784 was previosly studied
by Heidt \& Wagner (1996) for two different observing runs,
and for time lags greater than 1 day they report a slope of 0.58$\pm$0.13
practically coincident with our result.

In Fig. 9 we also report the SF for the radio light curve at 8.4 GHz, calculated
on the flux densities in Jy, with the same binning of the optical one, again
without any normalization.
The points are more scatted than the optical ones but the mean slope is
not significantly different (0.51$\pm$0.18).
No indication of periodicity is evident also in the radio data.

\section{ Optical energy distribution }

Besides the R band, most of the times also B, V and I band observations
were secured, to allow the determination of the broad band shape of the
energy distribution. Plots of the color indices B$-$V, V$-$R and R$-$I
as a function of time are reported in Fig. 10.
Their average values are B$-$V=0.59$\pm$0.05, V$-$R=0.49$\pm$0.04 and
R$-$I=0.63$\pm$0.03, where the quoted errors are 1 $\sigma$ deviations of
the data set.
Large systematic changes of the source colour are not evident, in
particular during the three large luminosity variations episodes
(see Figs. 3 to 5).

Several BL Lac objects have optical spectra well reproduced by a
power law ($F_{\nu}=A \nu^{\alpha}$). 
To derive the spectral index it is necessary to correct the observed fluxes for
the foreground absorption, which is expected to be not very large, given the
relatively low (b=29 degrees) galactic latitude of this source.
We evaluated the absorption from literature N$_H$ values
(3.90 $\times$ 10$^{20}$, Ciliegi et al. 1995; 3.70 $\times$ 10$^{20}$,
Murphy et al. 1996), and adopting the ratio N$_H$/E(B$-$V) = 5.2 $\times$
10$^{21}$ cm$^{-2}$mag$^{-1}$ from Shull \& Van Steenberg (1985).
The resulting colour excess E(B$-$V)=0.075 was derived, and the corresponding
extinctions in the B, V, R, I bands computed assuming the curve by Schlegel
et al. (1998). Conversion from magnitudes to fluxes was made according to Mead
et al. (1990). Any contribution from the host galaxy is likely negligible, given
that the CaII break is undetectable in the published spectra (Lawrence et al. 1996;
Rector \& Stocke 2001).

We have 85 4-band simultaneous observations of this source, useful to derive 
the spectral index.
A fit of log($F_{\nu}$) vs log($\nu$) with a straight line of slope $\alpha$
gives a satisfactory result against a $\chi ^2$ test for nearly all our
simultaneous 4-bands observations, supporting the power law modelling of the
spectral energy distribution.
We report in Table 4 the JD, R, $\sigma (R)$, $\alpha$, $\sigma(\alpha)$ and 
$\chi ^2$ for each day. Consistent results are obtained using the
larger set (134) of good quality V, R, I simultaneous observations.
Our average value (-1.54, with a dispersion of 0.12) is in good agreement
with the result of the only spectrophotometric observation of
S5 1803+784 available in the literature (-1.48; Lawrence et al., 1996).
A plot of the spectral index {\it vs} the R magnitude is shown in Fig. 11.
The linear correlation coefficient was found equal to 0.42, which means a
probability less than 10$^{-3}$ that the correlation is merely due to chance.
From this plot it is apparent however that at relatively large changes of the
luminosity do not correspond large variations of the spectral
slope, at variance with other BL Lac objects, like ON 231 (Massaro et
al. 1999) or BL Lacertae (Nesci et al. 1998).

\section{Discussion}

In this paper we presented the first flux calibrated,
long term variability study of the BL Lac object
S5 1803+784 in the optical frequency range, spanning more than five years
since Spring 1996.
Furthermore, we made a comparison with the behaviour of this source
in the same period in the radio band at 8.4 GHz.

The main results can be summarized as follows:

a) the source showed an optical luminosity variation with an amplitude
of a factor $\sim$ 20;

b) the general behaviour was characterized by an increase of the mean
flux level for at least 1150 days during which we observed with three large
flares, followed by a decrease in about 400 days down to a much lower state;

c) there is no evidence of periodicity in the light curve on the time scales
(several months) well sampled by our dataset;

d) the three major flares were characterized by comparable rise and decay times of 
about 20 days, corresponding to an (unbeamed)
dimension of the emitting region $R \simeq c \Delta t \simeq$ 5 $\times$ 10$^{16}$ cm;

e) changes of the source luminosity are not related to strong systematic trends
of the spectral slope $\alpha$.
A correlation test provided evidence for an increase of the slope in fainter states:
however the $\alpha$ values corresponding to the highest and lowest luminosity
differ only by $\sim$0.1 while larger differences were occasionally found in 
intermediate states;

f) the source showed variability in the radio band, with
maximum to minimum difference of 1 Jy and a mean flux density
value of $\sim$3 Jy at 8.4 GHz. 

The slopes of the optical spectrum, derived in Section 3,
indicate that the synchrotron emission peak of the SED should be in the IR
range. The only JHK photometric data available in the literature
are those by Heckman et al. (1983) taken in 1980-81
and give a much flatter slope (about -0.8). 
The IRAS data (Impey \& Neugebauer 1988) give a substantially
flat energy distribution suggesting that the SED should peak between 10 and 100
$\mu$m. 

To derive a more complete SED we retrieved from the ISOCAM archive the four
available calibrated images, taken in April 1996 but still unpublished, and 
made a simple aperture photometry.
In the X ray band, the source was lately observed by BeppoSAX on 28-Sep-1998
(Padovani et al. in preparation). The
2-10 keV spectrum had an energy spectral slope $\alpha _x$=-0.45,
significantly flatter than the optical one (-1.7), indicating that the X-ray
emission was completely dominated by the Inverse Compton component.
Our nearest optical data are on 20-Sep and 11-Oct
giving the source respectively at R=15.3 and R=15.7 (see Table 3).

Using these data, although not simultaneous, we constructed the SED, given in
Fig. 12, which improves on that given by Giommi et et. (1996).
For our optical and radio monitoring we plotted only the minimum and 
maximum values, to give a feeling of the variability of the source.
For the JHK bands we plotted our data, together with the simultaneous optical ones.
The R value form our monitoring simultaneous to the ISOCAM observation is also
plotted. The general impression of a peak of the SED in the infrared range is
confirmed from Fig. 12. 

It is interesting to compare the luminosity and the behaviour of S5 1803+784
with that of other BL Lac sources at their maximum luminosity.
At the highest level (R=14), adopting $H_0$=65 km s$^{-1}$ Mpc$^{-1}$
and $q_0$=0.5, the absolute magnitude of the source is M$_R$=$-$28.7,
corresponding to an integrated optical luminosity of 3.6 10$^{46}$ erg s$^{-1}$.
ON 231, which reached the highest recorded flux (R=12.2) in Spring 1998
(Massaro et al. 1999), at a redshift $z=0.10$, reached an absolute luminosity 
M=$-$26.2, i.e. 2.5 mag weaker than S5 1803+784. BL Lac, during the large flare in
1997 (Bloom et al. 1997) reached R=12.3, corresponding to M=$-$25.2, i.e. 3.5 mag
fainter than S5 1803+784.
These different absolute luminosities may be due to an intrinsic different power or
to a different beaming factor. L\"ahteenm\"aki \& Valtaoja (1999)
estimated Doppler boosting factors $\delta$ from variations of brightness
temperature at 22 and 33 GHz and found values of 1.6 for ON 231, 3.9 for BL Lac and
6.5 for S5 1803+784. The higher beaming could be the reason of the apparent
higher luminosity of our source: however, in this case, the higher time contraction 
would imply large amplitude flux variations on quite short time scales, which have not
been detected in our observation, at variance with ON 231 and BL Lac (Nesci et al.
1998).

We can also estimate the unbeamed luminosity using the factor $\delta$
given above, and from the relation

\begin{equation}
 L_{obs}= \delta^{(3-\alpha)} L_{unb},
\end{equation}

\medskip

\noindent
we obtain an intrinsic optical
luminosity of 7.8 10$^{42}$ erg/s, which is quite reasonable (Cavaliere \&
Malquori 1999).

A simple radiative cooling mechanism of freshly accelerated electrons cannot explain
the color behaviour of the bright flares of S5 1803+784, in particular that on JD 711:
in fact one would expect a
significant steepening of the spectral slope in the dimming phase, like
that clearly detected in the flares of ON 231 (Massaro et al. 1999) which is not the
case. A possible explanation would be that the observed spectral shape is the result of
an equilibrium condition between acceleration and radiation processes, while
the total number of particles involved, which determine the flux level, changes
with time. In this case
however one should invoke that the injection and leakage time scales must be
similar.
An alternative possibility is that the flares are originated by a change of
$\delta$. Being the flux proportional to $\delta^{(3-\alpha)}$
a variation of $\Delta m$ magnitudes would require a variation
of $\delta$ according to the relation:

\begin{equation}
log(\delta_2/\delta_1)=0.4(m_1 - m_2)/(3-\alpha),
\end{equation}

\medskip

\noindent
which for $\alpha \sim -$1.5 implies a variation of $\delta$ by a factor 1.5
for $\Delta m$ = 2, which looks rather substantial.

\centerline{\bf ACKNOWLEDGEMENTS}
The criticisms of an anonimous referee has helped to improve the first version
of this paper.
Part of this work was supported by the Italian Ministry for University and
Scientific Research (MURST) with grant Cofin-98-02-32 and Cofin 2001/028773.
This research has also made use of the NASA/IPAC Extragalactic Database (NED) which is 
operated by the Jet Propulsion Laboratory, California Institute of
Technology, under contract with the National Aeronautics and Space Administration.

\clearpage


\begin{figure}
\caption
{ The finding chart of S5 1803+784 from the POSS-I 103aE plate.
Reference stars are marked according to Table 1; the blazar is marked with S.
 \label{f1} }
\end{figure}

\begin{figure}
\caption
{The R$_C$ light curve of S5 1803+784 from Apr 1996 to Jan 2002.
 \label{f2}}
\end{figure}

\begin{figure}
\caption
{First flare. The R light curve (filled boxes) and the V$-$I colour index
(crosses with error bars).
The colour index scale is on the right side, the R magnitude scale is
on the left. No appreciable variation of the V$-$I colour index is present.
 \label {f3}}
\end{figure}

\begin{figure}
\caption
{Second flare. The R light curve (filled boxes) and the V$-$I colour index
(crosses with error bars). Scales as in Fig.3
 \label{f4} }
\end{figure}

\begin{figure}
\caption
{Third flare. The R light curve (filled boxes) and the V$-$I colour index
(crosses with error bars).  Scales as in Fig. 3.
Marginal evidence for a redder colour in the decreasing branch is
present. \label{f5}}
\end{figure}

\begin{figure}
\caption
{The radio light curve of S5 1803+784 at 8.4 GHz since 1996
(upper panel), and the R light
curve in linear scale (mJy) not corrected for reddening (lower panel).
 \label{f6} }
 \end{figure}

\begin{figure}
\caption
{Power spectra of the light curve (upper panel) and of the sampling
window (lower panel). The peaks corresponding to 218 and 105 days are
indicated.
\label{f7}}
\end{figure}

\begin{figure}
\caption
{The Jurkevich test for 5 days step periods. \label{f8}}
 \end{figure}

\begin{figure}
\caption
{The structure function for the optical data, averaged over log($\Delta
t_{days}$)=0.1 (filled squares) and of the radio data averaged in the same
way (open circles). No normalization has ben applied to the data.
\label{f9}}
\end{figure}

\begin{figure}
\caption
{B$-$V, V$-$R and R$-$I colour indices plotted as a function
of time. Some data with large errors have been omitted. \label{f10}}
\end{figure}

\begin{figure}
\caption
{The 4-band spectral index as a function of the R magnitude.
The linear correlation coefficient is 0.42. \label{f11}  }
\end{figure}

\begin{figure}
\caption
{The spectral energy distribution of S5 1803+784 derived from literature and our
measurements: NED data (radio and IRAS), open squares; radio data
(max and min, this paper), filled squares;
ISO and simultaneous optical data, filled triangles; simultaneous NIR and optical
data (this paper), open triangles; optical data (max and min, this paper),
filled circles; BeppoSAX x-ray data, crosses. \label{f12}  }
\end{figure}

\clearpage

\begin{deluxetable}{cllll}
\tablecaption{Magnitudes of comparison stars for S5 1803+784. \label{tbl-1}}
\tablewidth{0pt}
\tablehead{
\colhead{Star\tablenotemark{a}} & \colhead{B} & \colhead{V} & \colhead{R} & \colhead{I}
}
\startdata
A& 15.39$\pm$.05 (7)&14.54$\pm$.03(8) & 14.06$\pm$.03(10)& 13.62$\pm$.03(8) \\
B  & 15.97$\pm$.01 &15.30$\pm$.01 & 14.92$\pm$.01& 14.53$\pm$.01 \\
C  & 16.37$\pm$.01 &15.73$\pm$.01 & 15.36$\pm$.01& 14.97$\pm$.01 \\
D  &               &16.58$\pm$.01 & 16.08$\pm$.01& 15.57$\pm$.01 \\
E  &               &16.49$\pm$.01 & 16.10$\pm$.01& 15.70$\pm$.01 \\
\enddata
\tablenotetext{a}{Errors for star A include the zero point uncertainty, while
for the others only the intercalibration uncertainty is given}

\end{deluxetable}

\clearpage

\begin{deluxetable}{crrcc}
\tablecaption{Photographic archive magnitudes of S5 1803+784. \label{tbl-2}}
\tablewidth{0pt}
\tablehead{
\colhead{Band} & \colhead{mag} & \colhead{date} & \colhead{Survey} &
\colhead{emulsion}
}
\startdata
 R &14.5 &11-08-1953& POSS-I& 103aE \\
 V &15.5 &16-05-1983& Quick Blue& IIaD  \\
 R &15.5 &21-08-1993& POSS-II& IIIaF \\
 B &16.6 &12-09-1993& POSS-II& IIIaJ \\
\enddata
\end{deluxetable}

\clearpage

\begin{deluxetable}{rrrrrrrrrr}
\tabletypesize{\scriptsize}
\tablecaption{B,V,R,I magnitudes of S5 1803+784. \label{tbl-3}}
\tablewidth{0pt}
\tablehead{
\colhead{JD\tablenotemark{a}} & \colhead{B} & \colhead{$\sigma$B} & \colhead{V} &
\colhead{$\sigma$V}
& \colhead{R} & \colhead{$\sigma$R} & \colhead{I} & \colhead{$\sigma$I} &
\colhead{Tel}
}
\startdata
     193.4382 &  0.00 &  0.00 & 16.53 &  0.05 & 15.96 &  0.03 & 15.38 &  0.03 & VA \\
     222.4194 &  0.00 &  0.00 &  0.00 &  0.00 & 15.93 &  0.03 &  0.00 &  0.00 & VA \\
     246.4236 &  0.00 &  0.00 &  0.00 &  0.00 & 15.76 &  0.02 &  0.00 &  0.00 & MP \\
     272.4507 &  0.00 &  0.00 & 16.87 &  0.03 &  0.00 &  0.00 & 15.73 &  0.03 & VA \\
     277.5292 &  0.00 &  0.00 & 17.03 &  0.05 & 16.55 &  0.05 & 15.90 &  0.03 & VA \\
     354.4250 &  0.00 &  0.00 & 16.28 &  0.06 & 15.82 &  0.03 &  0.00 &  0.00 & VA \\
     357.3854 &  0.00 &  0.00 &  0.00 &  0.00 & 15.87 &  0.02 &  0.00 &  0.00 & VA \\
     375.3771 &  0.00 &  0.00 & 16.67 &  0.03 & 16.21 &  0.03 &  0.00 &  0.00 & VA \\
     377.3847 &  0.00 &  0.00 & 16.67 &  0.03 & 16.24 &  0.03 & 15.55 &  0.05 & VA \\
     378.4056 &  0.00 &  0.00 & 16.64 &  0.03 & 16.19 &  0.03 & 15.53 &  0.03 & VA \\
     381.2625 &  0.00 &  0.00 & 16.60 &  0.03 & 16.19 &  0.03 &  0.00 &  0.00 & VA \\
     390.3125 &  0.00 &  0.00 & 16.61 &  0.05 & 16.16 &  0.05 &  0.00 &  0.00 & VA \\
     391.2313 &  0.00 &  0.00 & 16.60 &  0.03 & 16.18 &  0.05 &  0.00 &  0.00 & VA \\
     415.3014 &  0.00 &  0.00 & 16.27 &  0.03 & 15.82 &  0.04 & 15.18 &  0.03 & VA \\
     421.2299 &  0.00 &  0.00 & 16.57 &  0.03 & 16.11 &  0.03 &  0.00 &  0.00 & VA \\
     464.2500 & 16.67 &  0.03 & 16.16 &  0.03 & 15.71 &  0.03 & 15.04 &  0.03 & VA \\
     476.3542 &  0.00 &  0.00 & 15.86 &  0.03 & 15.43 &  0.03 & 14.86 &  0.03 & VA \\
     487.5382 &  0.00 &  0.00 & 15.98 &  0.03 & 15.48 &  0.03 & 14.83 &  0.03 & VA \\
     488.6021 &  0.00 &  0.00 & 16.01 &  0.03 & 15.52 &  0.04 & 14.90 &  0.03 & VA \\
     517.4750 &  0.00 &  0.00 & 16.40 &  0.03 & 15.94 &  0.04 & 15.24 &  0.03 & VA \\
     521.4507 &  0.00 &  0.00 &  0.00 &  0.00 & 15.64 &  0.04 & 15.12 &  0.03 & VA \\
     549.4910 & 16.93 &  0.06 & 16.34 &  0.03 & 15.85 &  0.03 & 15.23 &  0.03 & VA \\
     572.4729 &  0.00 &  0.00 & 15.57 &  0.03 & 15.02 &  0.03 & 14.38 &  0.03 & VA \\
     598.5042 &  0.00 &  0.00 & 15.98 &  0.03 & 15.50 &  0.03 &  0.00 &  0.00 & VA \\
     628.4424 &  0.00 &  0.00 &  0.00 &  0.00 & 15.67 &  0.03 &  0.00 &  0.00 & VA \\
     639.4271 &  0.00 &  0.00 & 16.49 &  0.03 & 16.00 &  0.03 & 15.29 &  0.03 & VA \\
     669.3965 &  0.00 &  0.00 & 16.42 &  0.03 & 15.94 &  0.03 &  0.00 &  0.00 & VA \\
     711.3493 & 15.44 &  0.05 & 14.90 &  0.03 & 14.43 &  0.02 & 13.76 &  0.03 & VA \\
     712.2597 & 15.61 &  0.05 & 15.09 &  0.03 & 14.54 &  0.03 & 13.87 &  0.03 & VA \\
     717.2931 & 15.65 &  0.03 & 15.03 &  0.04 & 14.54 &  0.03 & 13.92 &  0.03 & VA \\
     719.2528 & 15.84 &  0.03 & 15.30 &  0.03 & 14.77 &  0.03 & 14.11 &  0.03 & VA \\
     721.3368 & 15.91 &  0.04 & 15.38 &  0.03 & 14.82 &  0.03 & 14.19 &  0.03 & VA \\
     723.2437 & 16.02 &  0.05 & 15.57 &  0.03 & 15.06 &  0.03 & 14.43 &  0.03 & VA \\
     723.4479 & 15.97 &  0.04 & 15.51 &  0.03 & 14.99 &  0.03 & 14.37 &  0.03 & VA \\
     725.3299 &  0.00 &  0.00 &  0.00 &  0.00 & 15.15 &  0.02 &  0.00 &  0.00 & VA \\
     727.3264 & 16.77 &  0.03 & 16.04 &  0.03 & 15.51 &  0.03 & 14.88 &  0.03 & VA \\
     731.3243 & 16.93 &  0.04 & 16.26 &  0.04 & 15.78 &  0.03 & 15.13 &  0.03 & MP \\
     739.2917 &  0.00 &  0.00 & 16.30 &  0.04 & 15.82 &  0.03 & 15.18 &  0.03 & VA \\
     742.2708 &  0.00 &  0.00 &  0.00 &  0.00 & 15.64 &  0.03 &  0.00 &  0.00 & VA \\
     744.2792 & 16.64 &  0.04 & 16.06 &  0.03 & 15.51 &  0.03 & 14.88 &  0.03 & VA \\
     747.2549 & 16.46 &  0.03 & 15.90 &  0.03 & 15.40 &  0.04 & 14.78 &  0.03 & VA \\
     748.3292 & 16.46 &  0.03 & 15.93 &  0.03 & 15.39 &  0.03 & 14.75 &  0.03 & VA \\
     756.2889 &  0.00 &  0.00 & 15.92 &  0.03 & 15.44 &  0.03 & 14.82 &  0.03 & VA \\
     781.2681 & 16.99 &  0.09 & 16.28 &  0.02 & 15.70 &  0.03 & 15.09 &  0.03 & MP \\
     791.2174 &  0.00 &  0.00 &  0.00 &  0.00 & 15.27 &  0.03 &  0.00 &  0.00 & VA \\
     823.2160 & 16.12 &  0.05 & 15.62 &  0.03 & 15.11 &  0.03 & 14.45 &  0.03 & VA \\
     825.2132 &  0.00 &  0.00 & 15.55 &  0.03 & 15.04 &  0.03 & 14.37 &  0.03 & VA \\
     832.2194 &  0.00 &  0.00 & 15.89 &  0.04 & 15.35 &  0.02 & 14.75 &  0.02 & VA \\
     838.2549 &  0.00 &  0.00 & 16.13 &  0.04 & 15.55 &  0.02 & 14.92 &  0.02 & VA \\
     840.2556 &  0.00 &  0.00 & 16.15 &  0.03 & 15.69 &  0.01 & 15.02 &  0.02 & VA \\
     842.2257 &  0.00 &  0.00 & 16.14 &  0.03 & 15.63 &  0.02 & 14.96 &  0.03 & VA \\
     843.2250 &  0.00 &  0.00 & 16.20 &  0.02 & 15.63 &  0.01 & 15.02 &  0.04 & VA \\
     858.2444 & 16.65 &  0.08 &  0.00 &  0.00 & 15.55 &  0.02 &  0.00 &  0.00 & VA \\
     860.2382 &  0.00 &  0.00 & 16.02 &  0.02 & 15.53 &  0.02 & 14.85 &  0.03 & VA \\
     862.2472 &  0.00 &  0.00 & 16.21 &  0.02 & 15.58 &  0.01 & 14.87 &  0.01 & VA \\
     863.2597 & 16.58 &  0.06 & 16.09 &  0.03 & 15.53 &  0.01 & 14.83 &  0.01 & VA \\
     865.5174 & 16.55 &  0.05 & 15.98 &  0.02 & 15.40 &  0.01 & 14.74 &  0.01 & VA \\
     871.5194 & 16.28 &  0.03 & 15.71 &  0.02 & 15.11 &  0.01 & 14.57 &  0.01 & VA \\
     872.4646 & 16.37 &  0.02 & 15.75 &  0.01 & 15.15 &  0.01 & 14.59 &  0.05 & VA \\
     891.4333 & 16.40 &  0.04 & 15.84 &  0.02 & 15.28 &  0.01 & 14.62 &  0.01 & VA \\
     895.5132 &  0.00 &  0.00 & 15.89 &  0.02 & 15.36 &  0.01 & 14.68 &  0.01 & VA \\
     900.3889 &  0.00 &  0.00 & 15.89 &  0.02 & 15.37 &  0.02 &  0.00 &  0.00 & VA \\
     901.4056 & 16.59 &  0.05 & 15.89 &  0.03 & 15.46 &  0.02 & 14.71 &  0.02 & VA \\
     907.4604 &  0.00 &  0.00 &  0.00 &  0.00 & 15.33 &  0.02 &  0.00 &  0.00 & VA \\
     924.5722 &  0.00 &  0.00 & 15.37 &  0.02 & 14.83 &  0.01 & 14.20 &  0.01 & VA \\
     928.4799 & 15.91 &  0.07 & 15.45 &  0.02 & 14.90 &  0.01 & 14.26 &  0.01 & VA \\
     942.3563 &  0.00 &  0.00 &  0.00 &  0.00 & 15.10 &  0.04 &  0.00 &  0.00 & VA \\
     947.4076 &  0.00 &  0.00 & 15.46 &  0.04 & 14.93 &  0.03 &  0.00 &  0.00 & VA \\
     950.4188 &  0.00 &  0.00 &  0.00 &  0.00 & 14.67 &  0.01 & 14.02 &  0.04 & VA \\
     953.4444 &  0.00 &  0.00 & 15.48 &  0.02 & 14.90 &  0.02 &  0.00 &  0.00 & VA \\
     955.4368 & 15.86 &  0.03 & 15.31 &  0.03 & 14.82 &  0.01 & 14.10 &  0.01 & VA \\
     956.4264 & 15.70 &  0.04 & 15.12 &  0.05 & 14.54 &  0.01 & 13.96 &  0.01 & VA \\
     966.3799 &  0.00 &  0.00 & 15.15 &  0.01 & 14.64 &  0.02 & 14.06 &  0.01 & VA \\
     970.4146 &  0.00 &  0.00 & 15.86 &  0.04 & 15.26 &  0.03 & 14.57 &  0.02 & VA \\
     983.4451 &  0.00 &  0.00 & 15.97 &  0.03 & 15.45 &  0.01 & 14.77 &  0.02 & VA \\
     985.4597 & 16.53 &  0.02 & 15.91 &  0.01 & 15.38 &  0.02 & 14.73 &  0.01 & VA \\
     988.4215 &  0.00 &  0.00 & 15.86 &  0.05 & 15.39 &  0.02 & 14.70 &  0.02 & VA \\
     993.4562 & 16.03 &  0.01 & 15.48 &  0.01 & 14.94 &  0.01 & 14.34 &  0.01 & VA \\
     997.5181 &  0.00 &  0.00 & 15.17 &  0.03 & 14.71 &  0.01 & 14.11 &  0.01 & VA \\
    1001.4118 & 15.64 &  0.02 & 15.06 &  0.02 & 14.64 &  0.02 & 14.03 &  0.02 & VA \\
    1004.3958 & 15.79 &  0.02 & 15.28 &  0.02 & 14.78 &  0.02 & 14.23 &  0.02 & VA \\
    1005.3806 & 16.01 &  0.04 & 15.42 &  0.02 & 14.97 &  0.02 & 14.38 &  0.02 & VA \\
    1008.3944 & 16.07 &  0.02 & 15.52 &  0.04 & 15.11 &  0.02 & 14.47 &  0.03 & VA \\
    1011.4000 &  0.00 &  0.00 & 15.50 &  0.02 & 15.04 &  0.01 & 14.44 &  0.02 & VA \\
    1013.3979 & 15.98 &  0.03 & 15.39 &  0.02 & 14.94 &  0.01 & 14.41 &  0.02 & VA \\
    1015.3826 &  0.00 &  0.00 & 15.38 &  0.02 & 14.89 &  0.02 &  0.00 &  0.00 & VA \\
    1016.4271 &  0.00 &  0.00 & 15.35 &  0.04 & 14.87 &  0.03 & 14.25 &  0.03 & GR \\
    1018.3799 & 15.91 &  0.02 & 15.38 &  0.02 & 14.91 &  0.02 & 14.32 &  0.03 & VA \\
    1019.4174 & 15.89 &  0.05 & 15.37 &  0.03 & 14.91 &  0.03 & 14.29 &  0.03 & GR \\
    1021.3382 & 16.02 &  0.02 & 15.41 &  0.01 & 14.96 &  0.01 & 14.39 &  0.02 & VA \\
    1025.4611 & 15.89 &  0.04 & 15.35 &  0.03 & 14.88 &  0.03 & 14.28 &  0.04 & GR \\
    1026.3556 & 15.89 &  0.02 & 15.23 &  0.01 & 14.79 &  0.01 & 14.24 &  0.01 & VA \\
    1037.4639 & 16.09 &  0.10 & 15.55 &  0.08 & 14.97 &  0.04 & 14.26 &  0.05 & GR \\
    1039.4326 &  0.00 &  0.00 & 15.62 &  0.09 & 15.01 &  0.03 & 14.34 &  0.05 & GR \\
    1040.3285 & 16.05 &  0.03 & 15.42 &  0.02 & 14.95 &  0.02 & 14.34 &  0.02 & VA \\
    1041.4229 &  0.00 &  0.00 & 15.45 &  0.05 & 14.98 &  0.03 & 14.27 &  0.03 & GR \\
    1042.3438 & 15.94 &  0.04 & 15.37 &  0.03 & 14.90 &  0.02 & 14.23 &  0.01 & VA \\
    1043.3354 & 15.87 &  0.05 & 15.31 &  0.03 & 14.85 &  0.01 & 14.20 &  0.02 & VA \\
    1048.3194 & 15.92 &  0.03 & 15.32 &  0.01 & 14.86 &  0.01 & 14.23 &  0.01 & VA \\
    1050.3146 & 15.97 &  0.02 & 15.34 &  0.01 & 14.87 &  0.01 & 14.25 &  0.01 & VA \\
    1052.3271 & 15.91 &  0.03 & 15.27 &  0.02 & 14.81 &  0.01 & 14.18 &  0.01 & VA \\
    1054.4174 &  0.00 &  0.00 &  0.00 &  0.00 & 14.72 &  0.01 & 14.07 &  0.03 & VA \\
    1057.5313 & 15.90 &  0.02 & 15.22 &  0.01 & 14.75 &  0.01 & 14.10 &  0.01 & VA \\
    1058.3424 & 15.86 &  0.02 & 15.25 &  0.02 & 14.73 &  0.01 & 14.10 &  0.01 & VA \\
    1067.2944 & 16.25 &  0.05 & 15.58 &  0.00 & 15.10 &  0.01 & 14.47 &  0.01 & VA \\
    1072.2847 & 16.37 &  0.01 & 15.71 &  0.02 & 15.20 &  0.01 & 14.57 &  0.01 & VA \\
    1077.3354 & 16.45 &  0.02 & 15.80 &  0.02 & 15.31 &  0.02 & 14.68 &  0.02 & VA \\
    1098.2444 & 16.73 &  0.03 & 16.14 &  0.05 & 15.69 &  0.02 & 15.06 &  0.01 & VA \\
    1100.2979 &  0.00 &  0.00 & 16.27 &  0.01 & 15.77 &  0.01 & 15.15 &  0.01 & VA \\
    1101.2889 & 16.86 &  0.04 & 16.25 &  0.03 & 15.76 &  0.02 & 15.13 &  0.01 & VA \\
    1108.2667 &  0.00 &  0.00 & 16.18 &  0.03 & 15.70 &  0.03 & 15.08 &  0.03 & VA \\
    1109.2410 & 16.72 &  0.01 & 16.18 &  0.04 & 15.63 &  0.02 & 15.03 &  0.01 & VA \\
    1111.3681 &  0.00 &  0.00 & 15.82 &  0.01 & 15.40 &  0.01 & 14.78 &  0.01 & VA \\
    1114.2347 & 16.28 &  0.02 & 15.76 &  0.02 & 15.30 &  0.02 & 14.70 &  0.02 & VA \\
    1125.2090 & 15.39 &  0.03 & 14.82 &  0.02 & 14.38 &  0.01 & 13.80 &  0.01 & GR \\
    1126.2056 & 15.26 &  0.01 & 14.73 &  0.01 & 14.30 &  0.01 & 13.70 &  0.01 & VA \\
    1127.2854 & 15.15 &  0.03 & 14.66 &  0.02 & 14.18 &  0.02 & 13.63 &  0.02 & MP \\
    1133.2986 & 15.21 &  0.02 & 14.66 &  0.03 & 14.20 &  0.01 & 13.61 &  0.01 & VA \\
    1135.2083 & 15.45 &  0.08 & 14.88 &  0.05 & 14.37 &  0.03 & 13.76 &  0.03 & MP \\
    1135.2715 & 15.44 &  0.01 & 14.83 &  0.03 & 14.40 &  0.01 & 13.80 &  0.01 & VA \\
    1138.3042 & 15.64 &  0.02 & 15.01 &  0.01 & 14.58 &  0.04 & 13.89 &  0.03 & MP \\
    1151.2493 &  0.00 &  0.00 &  0.00 &  0.00 & 14.57 &  0.02 &  0.00 &  0.00 & VA \\
    1154.5431 & 15.86 &  0.01 & 15.19 &  0.03 & 14.70 &  0.01 & 14.10 &  0.01 & VA \\
    1156.2472 &  0.00 &  0.00 & 15.24 &  0.03 & 14.73 &  0.01 & 14.15 &  0.01 & VA \\
    1160.2583 & 15.82 &  0.03 & 15.14 &  0.03 & 14.70 &  0.03 & 14.07 &  0.01 & VA \\
    1164.2410 & 15.76 &  0.02 & 15.14 &  0.02 & 14.69 &  0.02 & 14.07 &  0.02 & VA \\
    1294.4354 &  0.00 &  0.00 & 15.72 &  0.07 &  0.00 &  0.00 &  0.00 &  0.00 & GR \\
    1320.5000 &  0.00 &  0.00 &  0.00 &  0.00 & 14.62 &  0.04 &  0.00 &  0.00 & GR \\
    1327.3444 & 15.74 &  0.03 & 15.12 &  0.02 & 14.62 &  0.03 & 14.01 &  0.03 & VA \\
    1338.4424 & 15.26 &  0.03 & 14.65 &  0.03 & 14.16 &  0.03 & 13.56 &  0.03 & GR \\
    1338.4979 & 15.23 &  0.05 & 14.65 &  0.04 & 14.17 &  0.02 & 13.56 &  0.02 & MP \\
    1339.3847 & 15.22 &  0.03 & 14.55 &  0.03 & 14.09 &  0.03 & 13.47 &  0.03 & GR \\
    1344.3701 & 15.11 &  0.03 & 14.43 &  0.03 & 14.04 &  0.02 & 13.33 &  0.02 & GR \\
    1350.4347 & 15.21 &  0.03 & 14.58 &  0.03 & 14.10 &  0.03 & 13.53 &  0.03 & GR \\
    1353.3701 & 15.06 &  0.03 &  0.00 &  0.00 & 14.05 &  0.03 & 13.33 &  0.03 & GR \\
    1373.4035 & 15.27 &  0.03 & 14.71 &  0.03 & 14.19 &  0.03 & 13.58 &  0.03 & GR \\
    1405.3611 & 16.42 &  0.08 &  0.00 &  0.00 &  0.00 &  0.00 & 14.62 &  0.06 & GR \\
    1414.3722 & 16.30 &  0.08 & 15.73 &  0.05 & 15.21 &  0.03 & 14.48 &  0.03 & GR \\
    1422.3549 & 16.70 &  0.04 & 16.03 &  0.04 &  0.00 &  0.00 & 14.76 &  0.04 & GR \\
    1424.3854 & 16.53 &  0.06 & 15.90 &  0.03 & 15.40 &  0.03 & 14.76 &  0.03 & GR \\
    1428.3278 & 16.66 &  0.04 & 15.99 &  0.01 & 15.46 &  0.01 & 14.82 &  0.01 & VA \\
    1433.4410 & 16.56 &  0.05 &  0.00 &  0.00 &  0.00 &  0.00 & 14.81 &  0.03 & GR \\
    1434.3847 & 16.53 &  0.06 & 15.90 &  0.04 &  0.00 &  0.00 & 14.75 &  0.03 & GR \\
    1435.4042 & 16.71 &  0.05 & 16.12 &  0.04 &  0.00 &  0.00 & 14.88 &  0.03 & GR \\
    1445.2660 & 17.04 &  0.01 & 16.44 &  0.02 & 15.90 &  0.02 & 15.25 &  0.02 & VA \\
    1456.3312 & 16.97 &  0.09 & 16.47 &  0.04 & 15.88 &  0.03 & 15.24 &  0.04 & GR \\
    1457.3306 & 16.89 &  0.01 & 16.32 &  0.02 & 15.79 &  0.02 & 15.11 &  0.01 & VA \\
    1480.2382 &  0.00 &  0.00 & 16.56 &  0.03 & 16.04 &  0.03 & 15.43 &  0.02 & VA \\
    1484.2361 &  0.00 &  0.00 & 16.67 &  0.09 & 16.03 &  0.05 & 15.35 &  0.05 & GR \\
    1491.2243 & 17.29 &  0.06 & 16.66 &  0.06 & 16.08 &  0.03 & 15.41 &  0.04 & GR \\
    1508.2389 & 16.71 &  0.02 & 16.11 &  0.02 & 15.55 &  0.02 & 14.93 &  0.03 & MP \\
    1510.3312 &  0.00 &  0.00 & 16.06 &  0.02 & 15.68 &  0.02 & 15.12 &  0.02 & MP \\
    1520.2917 & 16.71 &  0.02 & 16.07 &  0.02 & 15.63 &  0.02 & 15.06 &  0.01 & VA \\
    1546.2326 &  0.00 &  0.00 & 16.16 &  0.04 & 15.64 &  0.04 & 14.96 &  0.04 & GR \\
    1549.4319 &  0.00 &  0.00 & 16.12 &  0.02 & 15.63 &  0.02 & 14.98 &  0.02 & VA \\
    1556.2347 &  0.00 &  0.00 & 16.01 &  0.05 & 15.56 &  0.04 & 14.89 &  0.04 & GR \\
    1569.2410 & 16.58 &  0.05 & 15.95 &  0.03 & 15.57 &  0.03 &  0.00 &  0.00 & GR \\
    1578.3493 &  0.00 &  0.00 &  0.00 &  0.00 & 15.45 &  0.02 &  0.00 &  0.00 & VA \\
    1606.4215 & 16.53 &  0.02 & 15.95 &  0.02 & 15.42 &  0.02 & 14.76 &  0.02 & VA \\
    1613.4931 &  0.00 &  0.00 & 15.86 &  0.02 & 15.31 &  0.02 & 14.62 &  0.03 & VA \\
    1655.3590 &  0.00 &  0.00 &  0.00 &  0.00 & 16.82 &  0.30 &  0.00 &  0.00 & GR \\
    1688.5451 &  0.00 &  0.00 & 17.19 &  0.03 & 16.80 &  0.03 & 16.01 &  0.04 & VA \\
    1695.3903 & 17.71 &  0.05 &  0.00 &  0.00 & 16.64 &  0.02 & 15.97 &  0.02 & VA \\
    1706.3597 &  0.00 &  0.00 &  0.00 &  0.00 & 16.57 &  0.02 &  0.00 &  0.00 & VA \\
    1717.5007 & 17.41 &  0.03 & 16.70 &  0.02 & 16.19 &  0.02 & 15.52 &  0.02 & VA \\
    1718.3965 &  0.00 &  0.00 & 16.74 &  0.05 & 16.23 &  0.04 &  0.00 &  0.00 & GR \\
    1721.4389 &  0.00 &  0.00 & 16.71 &  0.05 & 16.16 &  0.05 &  0.00 &  0.00 & GR \\
    1722.5472 &  0.00 &  0.00 & 16.79 &  0.02 & 16.20 &  0.02 & 15.48 &  0.02 & MP \\
    1724.4493 &  0.00 &  0.00 & 16.86 &  0.05 & 16.24 &  0.02 & 15.51 &  0.02 & MP \\
    1747.3424 &  0.00 &  0.00 &  0.00 &  0.00 & 16.25 &  0.04 &  0.00 &  0.00 & VA \\
    1749.3236 &  0.00 &  0.00 &  0.00 &  0.00 & 16.23 &  0.02 &  0.00 &  0.00 & VA \\
    1751.3542 &  0.00 &  0.00 &  0.00 &  0.00 & 16.27 &  0.03 &  0.00 &  0.00 & VA \\
    1752.3479 &  0.00 &  0.00 &  0.00 &  0.00 & 16.28 &  0.03 &  0.00 &  0.00 & VA \\
    1754.3458 &  0.00 &  0.00 &  0.00 &  0.00 & 16.31 &  0.02 &  0.00 &  0.00 & VA \\
    1758.3458 &  0.00 &  0.00 &  0.00 &  0.00 & 16.15 &  0.04 &  0.00 &  0.00 & VA \\
    1765.3257 &  0.00 &  0.00 &  0.00 &  0.00 & 16.46 &  0.03 &  0.00 &  0.00 & VA \\
    1768.3118 &  0.00 &  0.00 &  0.00 &  0.00 & 16.45 &  0.04 &  0.00 &  0.00 & VA \\
    1789.3687 & 17.38 &  0.04 & 16.87 &  0.02 & 16.24 &  0.02 & 15.64 &  0.03 & MP \\
    1814.5125 &  0.00 &  0.00 &  0.00 &  0.00 & 16.78 &  0.03 &  0.00 &  0.00 & MP \\
    1838.2986 &  0.00 &  0.00 &  0.00 &  0.00 & 16.05 &  0.05 &  0.00 &  0.00 & VA \\
    1841.2667 &  0.00 &  0.00 &  0.00 &  0.00 & 16.04 &  0.03 & 15.36 &  0.02 & VA \\
    1910.2062 &  0.00 &  0.00 &  0.00 &  0.00 & 16.47 &  0.08 &  0.00 &  0.00 & GR \\
    2012.4583 &  0.00 & 0.00 &  0.00 & 0.00 & 17.00 & 0.05 &  0.00 & 0.00 & GR\\
    2083.5090 &  0.00 & 0.00 &  0.00 & 0.00 & 16.98 & 0.05 &  0.00 & 0.00 & VA\\
    2087.3868 &  0.00 & 0.00 &  0.00 & 0.00 & 17.02 & 0.05 &  0.00 & 0.00 & VA\\
    2093.4472 &  0.00 & 0.00 &  0.00 & 0.00 & 17.30 & 0.08 &  0.00 & 0.00 & VA\\
    2098.4278 &  0.00 & 0.00 &  0.00 & 0.00 & 17.20 & 0.08 &  0.00 & 0.00 & VA\\
    2102.4535 &  0.00 & 0.00 &  0.00 & 0.00 & 17.22 & 0.08 &  0.00 & 0.00 & VA\\
    2103.3951 &  0.00 & 0.00 &  0.00 & 0.00 & 17.23 & 0.08 &  0.00 & 0.00 & VA\\
    2112.4097 &  0.00 & 0.00 &  0.00 & 0.00 & 17.24 & 0.03 & 16.43 & 0.02 & GR\\
    2113.3764 &  0.00 & 0.00 & 17.65 & 0.08 & 17.19 & 0.08 & 16.52 & 0.05 & VA\\
    2115.4042 &  0.00 & 0.00 &  0.00 & 0.00 & 17.35 & 0.08 & 16.65 & 0.08 & VA\\
    2116.4111 &  0.00 & 0.00 &  0.00 & 0.00 &  0.00 & 0.00 & 16.43 & 0.08 & GR\\
    2117.4153 &  0.00 & 0.00 &  0.00 & 0.00 & 17.18 & 0.08 & 16.59 & 0.06 & VA\\
    2119.4056 &  0.00 & 0.00 &  0.00 & 0.00 & 17.22 & 0.06 & 16.65 & 0.06 & VA\\
    2120.3701 &  0.00 & 0.00 &  0.00 & 0.00 & 17.22 & 0.08 &  0.00 & 0.00 & VA\\
    2129.3722 &  0.00 & 0.00 &  0.00 & 0.00 & 17.12 & 0.08 & 16.52 & 0.05 & VA\\
    2141.4139 &  0.00 & 0.00 &  0.00 & 0.00 & 16.86 & 0.05 &  0.00 & 0.00 & VA\\
    2148.3146 &  0.00 & 0.00 &  0.00 & 0.00 & 16.78 & 0.05 &  0.00 & 0.00 & VA\\
    2159.3069 &  0.00 & 0.00 &  0.00 & 0.00 & 16.85 & 0.09 &  0.00 & 0.00 & GR\\
    2164.3736 &  0.00 & 0.00 &  0.00 & 0.00 & 16.74 & 0.02 &  0.00 & 0.00 & VA\\
    2165.3438 &  0.00 & 0.00 & 17.24 & 0.08 & 16.70 & 0.05 & 16.07 & 0.03 & VA\\
    2172.3306 &  0.00 & 0.00 &  0.00 & 0.00 & 16.83 & 0.05 &  0.00 & 0.00 & VA\\
    2174.3521 &  0.00 & 0.00 &  0.00 & 0.00 & 16.79 & 0.05 & 16.18 & 0.03 & VA\\
    2178.3076 &  0.00 & 0.00 &  0.00 & 0.00 & 16.69 & 0.02 & 16.02 & 0.03 & VA\\
    2181.3042 &  0.00 & 0.00 &  0.00 & 0.00 & 16.65 & 0.05 &  0.00 & 0.00 & VA\\
    2194.2715 & 17.39 & 0.08 & 16.76 & 0.05 & 16.29 & 0.03 & 15.67 & 0.03 & VA\\
    2195.3174 &  0.00 & 0.00 &  0.00 & 0.00 & 16.29 & 0.03 &  0.00 & 0.00 & VA\\
    2236.2097 & 17.69 & 0.03 &  0.00 & 0.00 & 16.55 & 0.03 & 15.88 & 0.03 & TE\\
    2286.2333 & 17.01 & 0.08 & 16.38 & 0.03 & 15.91 & 0.03 & 15.24 & 0.02 & VA\\
    2287.2556 & 17.08 & 0.08 & 16.44 & 0.03 & 15.93 & 0.03 & 15.27 & 0.02 & VA\\
\enddata
\tablenotetext{a}{JD-2,450,000}
\tablenotetext{b} {The observation on JD 2236 was obtained with the 70 cm telescope
of the Collurania-Teramo Observatory by one of us (RN)}
\end{deluxetable}

\clearpage

\begin{deluxetable}{ccccc}
\tablecaption{Near Infrared observations. \label{tbl-4}}
\tablewidth{0pt}
\tablehead{
\colhead{dd-mm-yyyy}&\colhead{J}&\colhead{H}&\colhead{K}&
\colhead{ $\alpha_{nir}$}
}
\startdata
23-08-2001 & 14.99 & 14.09 & 13.26 & 1.26$\pm$0.11 \\
24-08-2001 & 15.06 & 14.04 & 13.27 & 1.36$\pm$0.11 \\
26-08-2001 & 15.00 & 14.00 &       &               \\
\enddata
\end{deluxetable}

\clearpage

\begin{deluxetable}{cccccc}
\tablecaption{Optical spectral slope fits. \label{tbl-5}}
\tablewidth{0pt}
\tablehead{
\colhead{JD-2400000} & \colhead{R} & \colhead{$\alpha$} &
\colhead{$\sigma (\alpha)$} & \colhead{$\chi ^2$}
}
\startdata
    464.2&15.71&-1.378&0.063& 3.037\\
    549.5&15.85&-1.509&0.084& 0.060\\
    711.3&14.43&-1.508&0.079& 1.356\\
    712.3&14.54&-1.647&0.079& 2.529\\
    717.3&14.54&-1.538&0.064& 0.031\\
    719.3&14.77&-1.553&0.063& 2.169\\
    721.3&14.82&-1.578&0.072& 2.340\\
    723.2&15.06&-1.427&0.079& 3.299\\
    723.4&14.99&-1.402&0.072& 3.556\\
    727.3&15.51&-1.785&0.063& 1.751\\
    731.3&15.78&-1.632&0.074& 0.154\\
    744.3&15.51&-1.619&0.072& 0.998\\
    747.3&15.40&-1.469&0.064& 0.534\\
    748.3&15.39&-1.527&0.063& 2.307\\
    781.3&15.70&-1.740&0.093& 1.830\\
    823.2&15.11&-1.530&0.079& 2.322\\
    863.3&15.53&-1.730&0.084& 3.715\\
    865.5&15.40&-1.736&0.079& 1.805\\
    871.5&15.11&-1.550&0.063& 4.092\\
    872.5&15.15&-1.704&0.079& 2.906\\
    891.4&15.28&-1.664&0.072& 1.813\\
    901.4&15.46&-1.730&0.079& 2.407\\
    928.5&14.90&-1.594&0.088& 2.859\\
    955.4&14.82&-1.581&0.063& 3.455\\
    956.4&14.54&-1.570&0.075& 1.581\\
    985.5&15.38&-1.654&0.063& 0.254\\
    993.5&14.94&-1.528&0.079& 1.151\\
   1001.4&14.64&-1.338&0.063& 0.327\\
   1004.4&14.78&-1.295&0.063& 1.034\\
   1005.4&14.97&-1.375&0.072& 0.033\\
   1008.4&15.11&-1.320&0.064& 1.259\\
   1013.4&14.94&-1.251&0.084& 0.476\\
   1018.4&14.91&-1.341&0.072& 0.364\\
   1019.4&14.91&-1.378&0.079& 0.737\\
   1021.3&14.96&-1.365&0.072& 0.330\\
   1025.5&14.88&-1.361&0.083& 0.296\\
   1026.4&14.79&-1.373&0.072& 1.831\\
   1037.5&14.97&-1.782&0.152& 0.636\\
   1040.3&14.95&-1.502&0.063& 0.165\\
   1042.3&14.90&-1.521&0.072& 0.931\\
   1043.3&14.85&-1.471&0.079& 0.631\\
   1048.3&14.86&-1.469&0.063& 0.090\\
   1050.3&14.87&-1.516&0.063& 0.102\\
   1052.3&14.81&-1.527&0.063& 0.216\\
   1057.5&14.75&-1.632&0.063& 0.508\\
   1058.3&14.73&-1.593&0.063& 0.222\\
   1067.3&15.10&-1.581&0.079& 0.301\\
   1072.3&15.20&-1.631&0.079& 0.206\\
   1077.3&15.31&-1.581&0.079& 0.101\\
   1098.2&15.69&-1.448&0.084& 0.172\\
   1101.3&15.76&-1.541&0.072& 0.016\\
   1109.2&15.63&-1.517&0.082& 1.100\\
   1114.2&15.30&-1.327&0.072& 0.564\\
   1125.2&14.38&-1.317&0.063& 0.017\\
   1126.2&14.30&-1.269&0.063& 0.484\\
   1127.3&14.18&-1.230&0.063& 0.961\\
   1133.3&14.20&-1.339&0.063& 0.128\\
   1135.2&14.37&-1.501&0.103& 0.172\\
   1135.3&14.40&-1.385&0.063& 0.317\\
   1138.3&14.58&-1.552&0.064& 0.785\\
   1154.5&14.70&-1.562&0.072& 0.789\\
   1160.3&14.70&-1.549&0.063& 1.161\\
   1164.2&14.69&-1.465&0.063& 0.165\\
   1327.3&14.62&-1.542&0.063& 0.139\\
   1338.4&14.16&-1.495&0.063& 0.120\\
   1338.5&14.17&-1.453&0.082& 0.040\\
   1339.4&14.09&-1.556&0.063& 0.756\\
   1344.4&14.04&-1.573&0.063& 3.000\\
   1350.4&14.10&-1.462&0.063& 0.750\\
   1373.4&14.19&-1.491&0.063& 0.760\\
   1414.4&15.21&-1.787&0.103& 1.155\\
   1424.4&15.40&-1.600&0.084& 0.010\\
   1428.3&15.46&-1.705&0.072& 0.308\\
   1445.3&15.90&-1.698&0.095& 0.334\\
   1456.3&15.88&-1.698&0.116& 1.912\\
   1457.3&15.79&-1.727&0.093& 0.512\\
   1491.2&16.08&-1.795&0.120& 0.350\\
   1508.2&15.55&-1.645&0.072& 1.028\\
   1520.3&15.63&-1.363&0.079& 0.726\\
   1606.4&15.42&-1.661&0.084& 0.568\\
   1717.5&16.19&-1.750&0.089& 0.197\\
   1789.4&16.24&-1.710&0.088& 4.749\\
   2194.3&16.29&-1.600&0.103& 1.084\\
   2286.2&15.91&-1.671&0.091& 0.989\\
   2287.3&15.93&-1.740&0.091& 0.837\\
\enddata
\end{deluxetable}

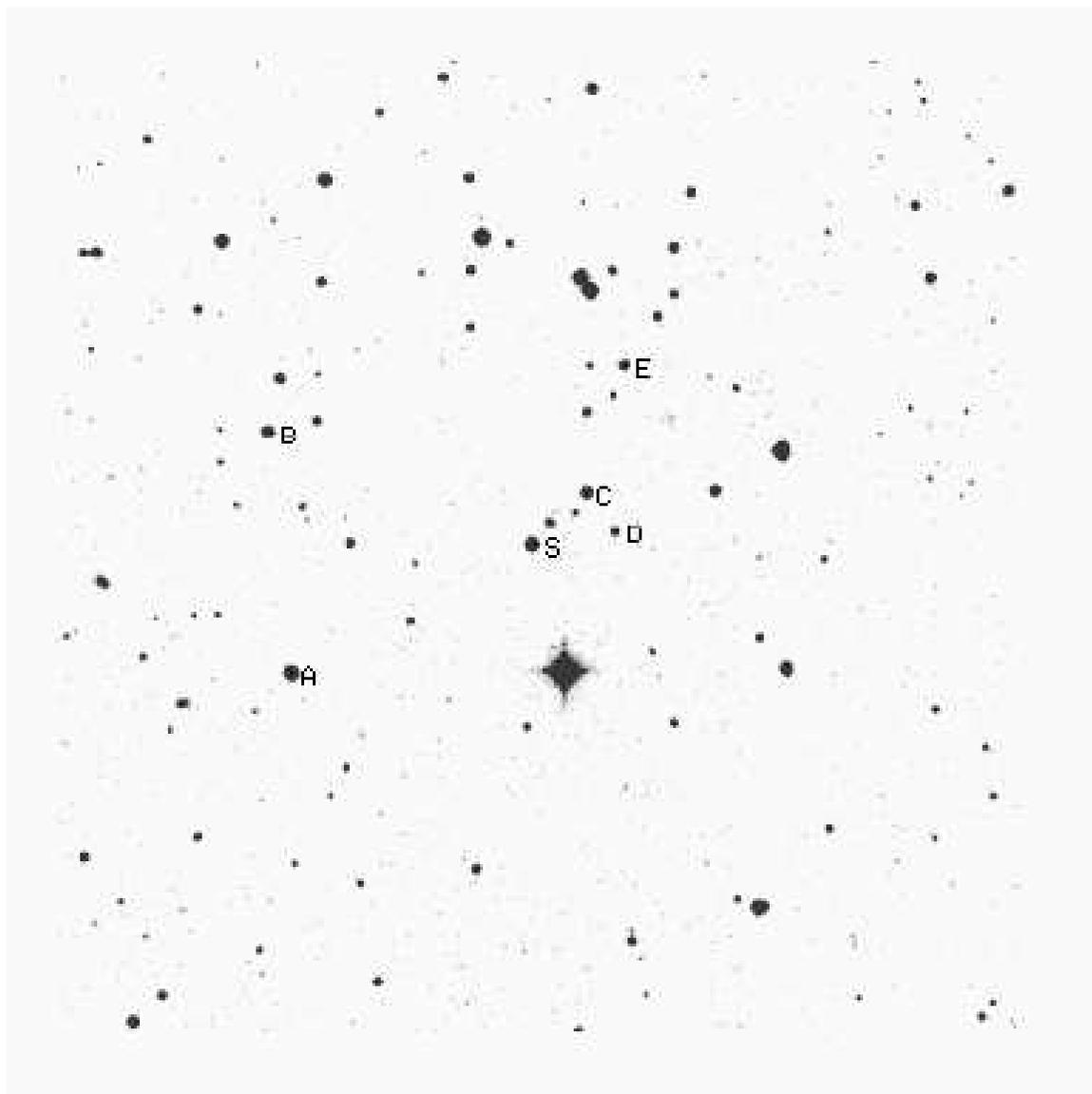
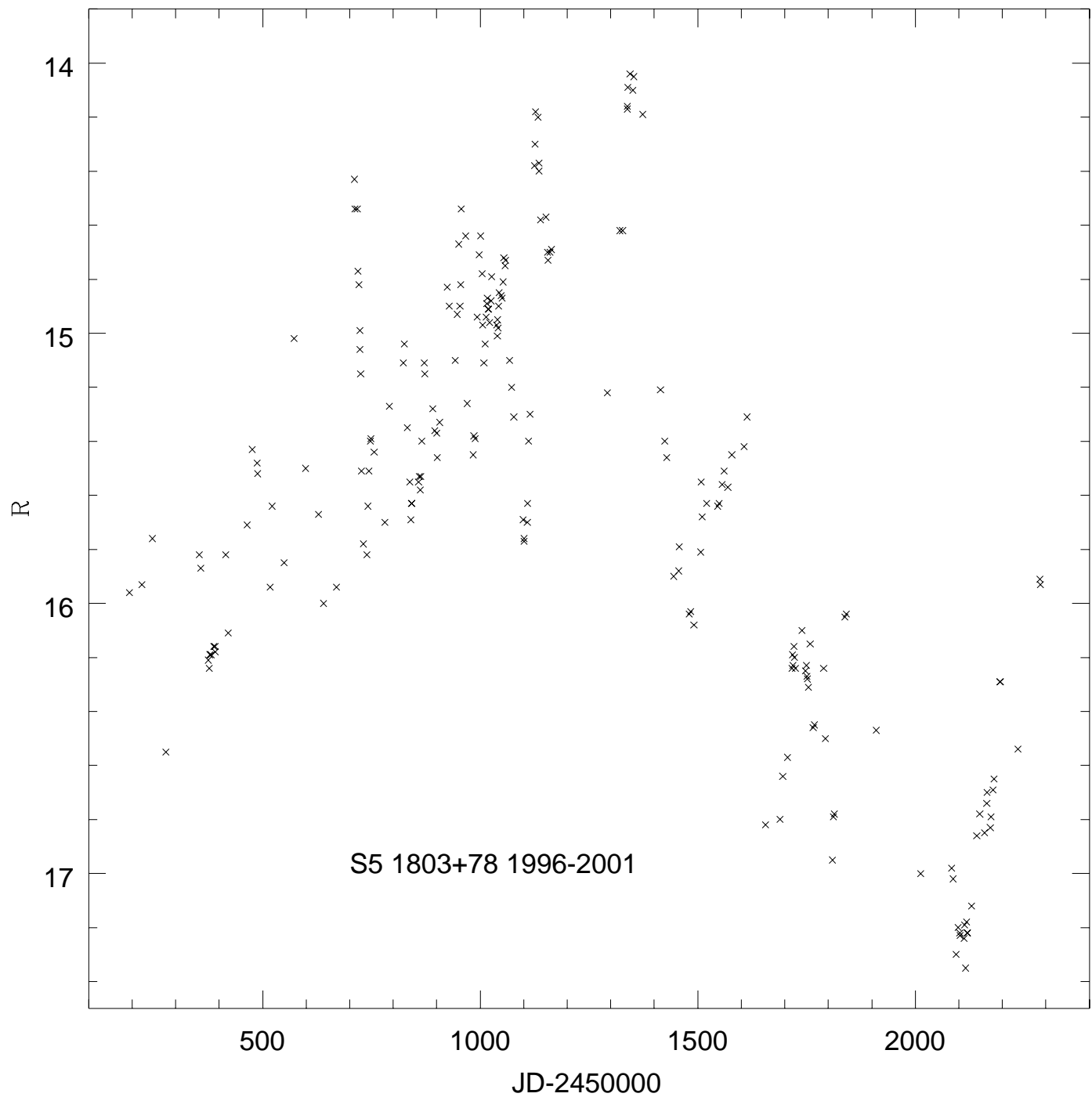
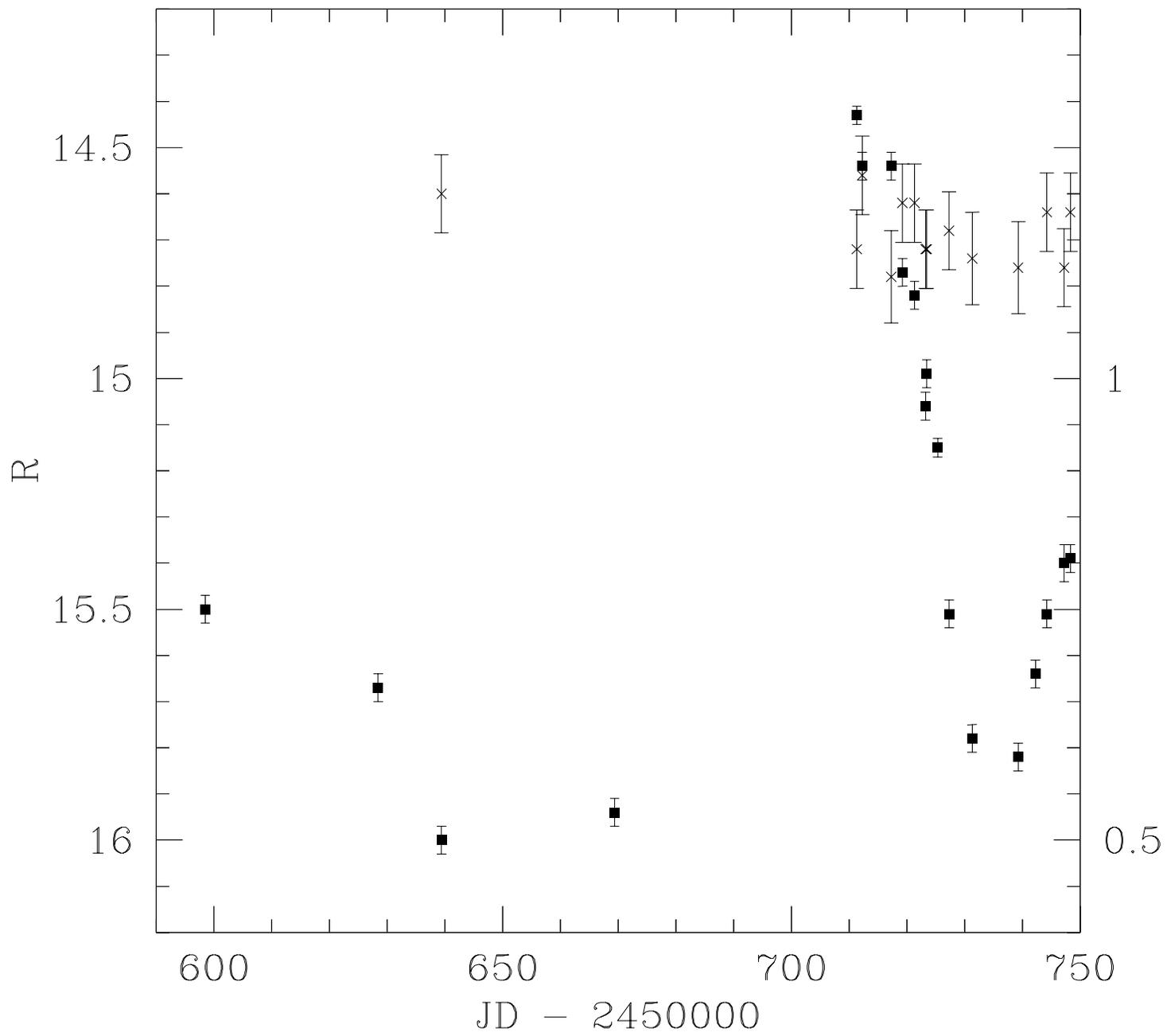
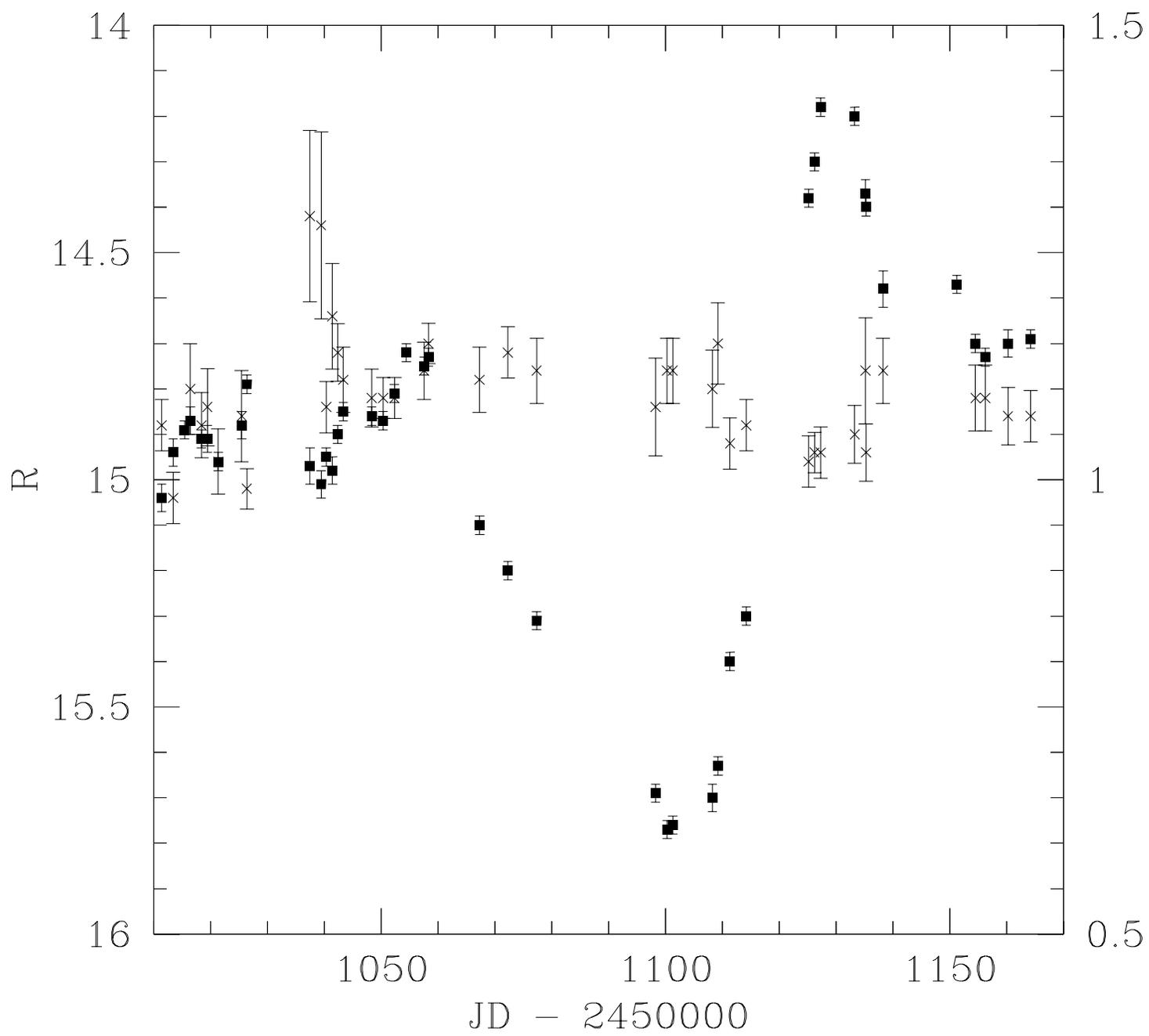
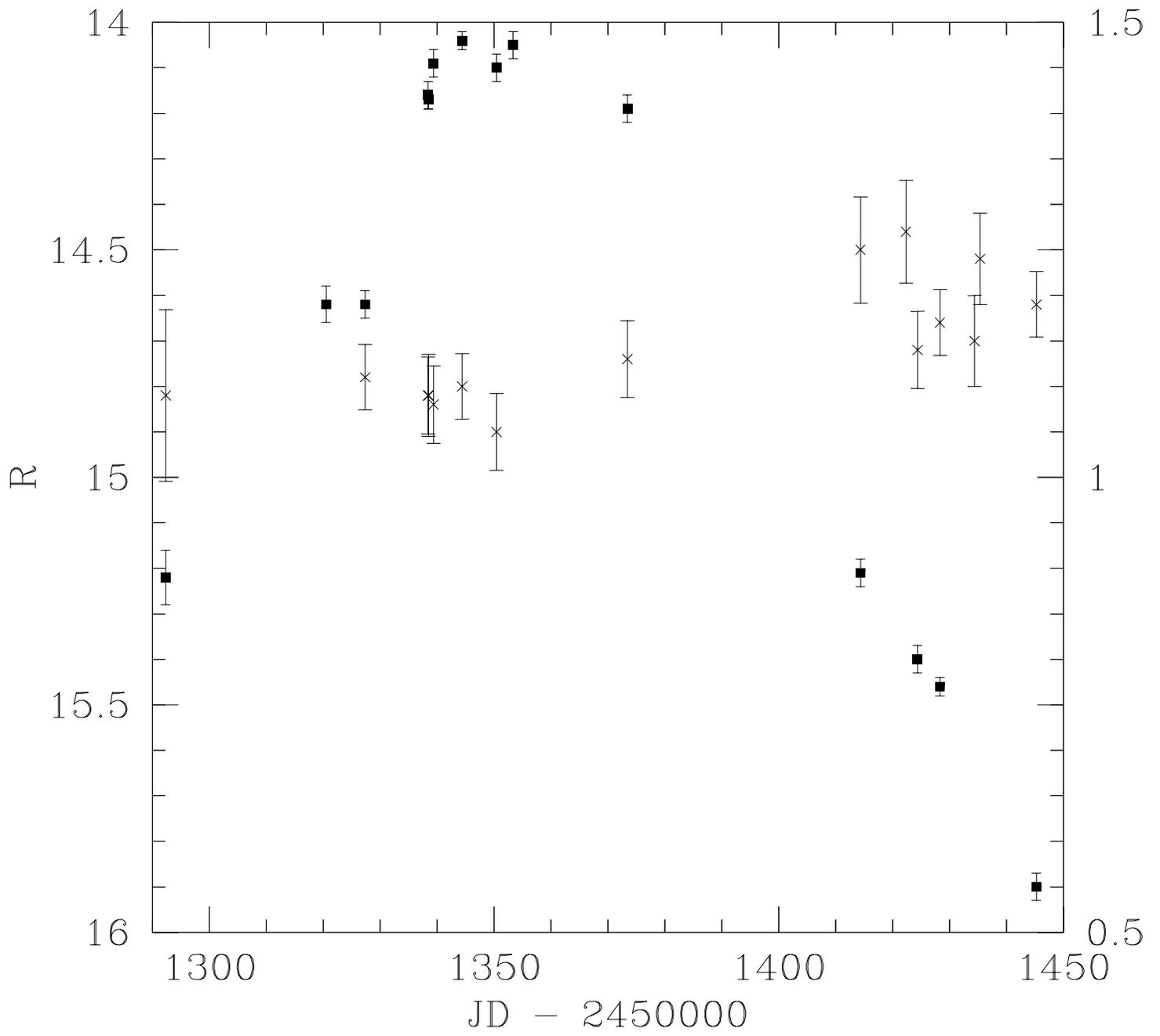
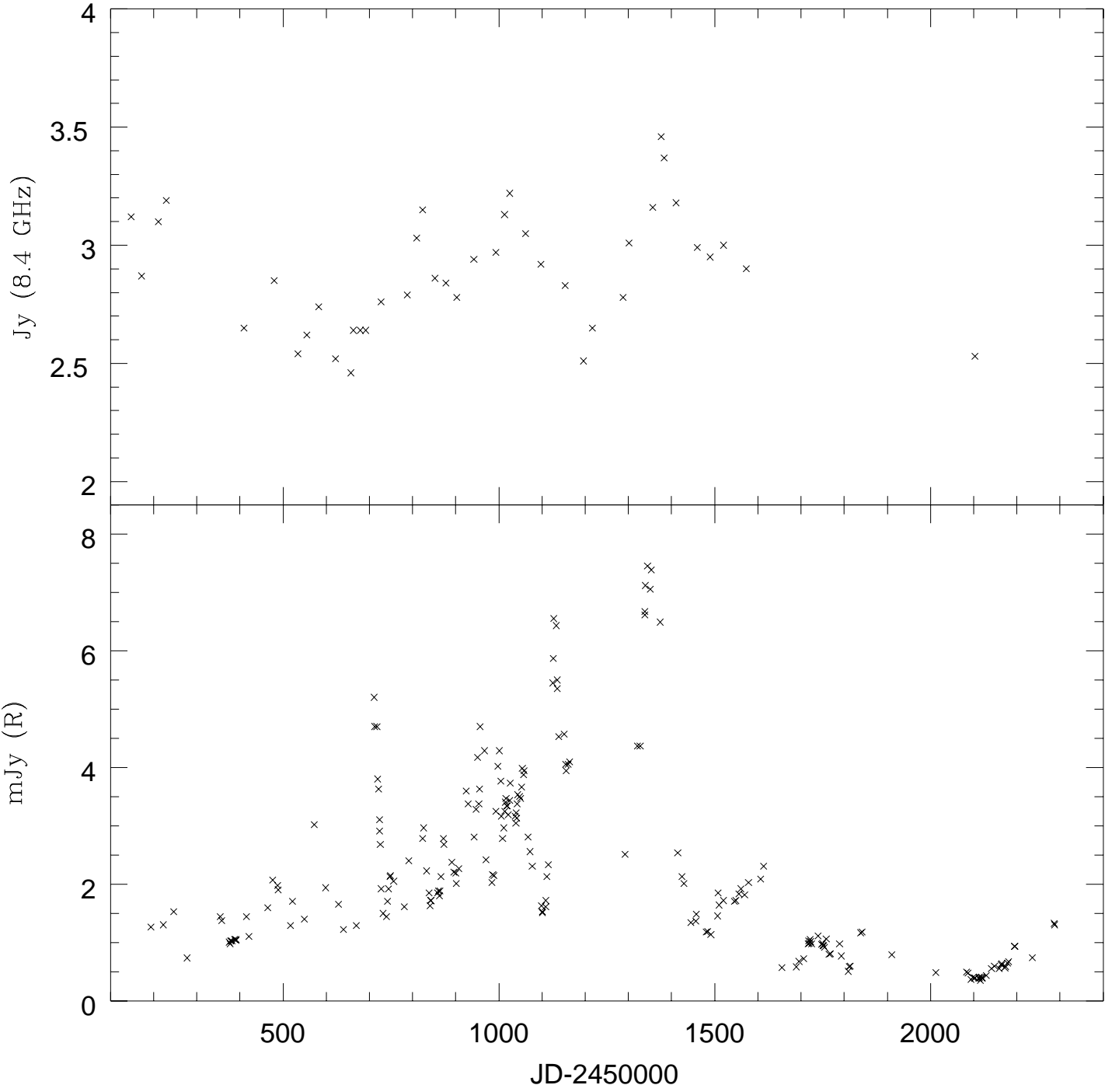
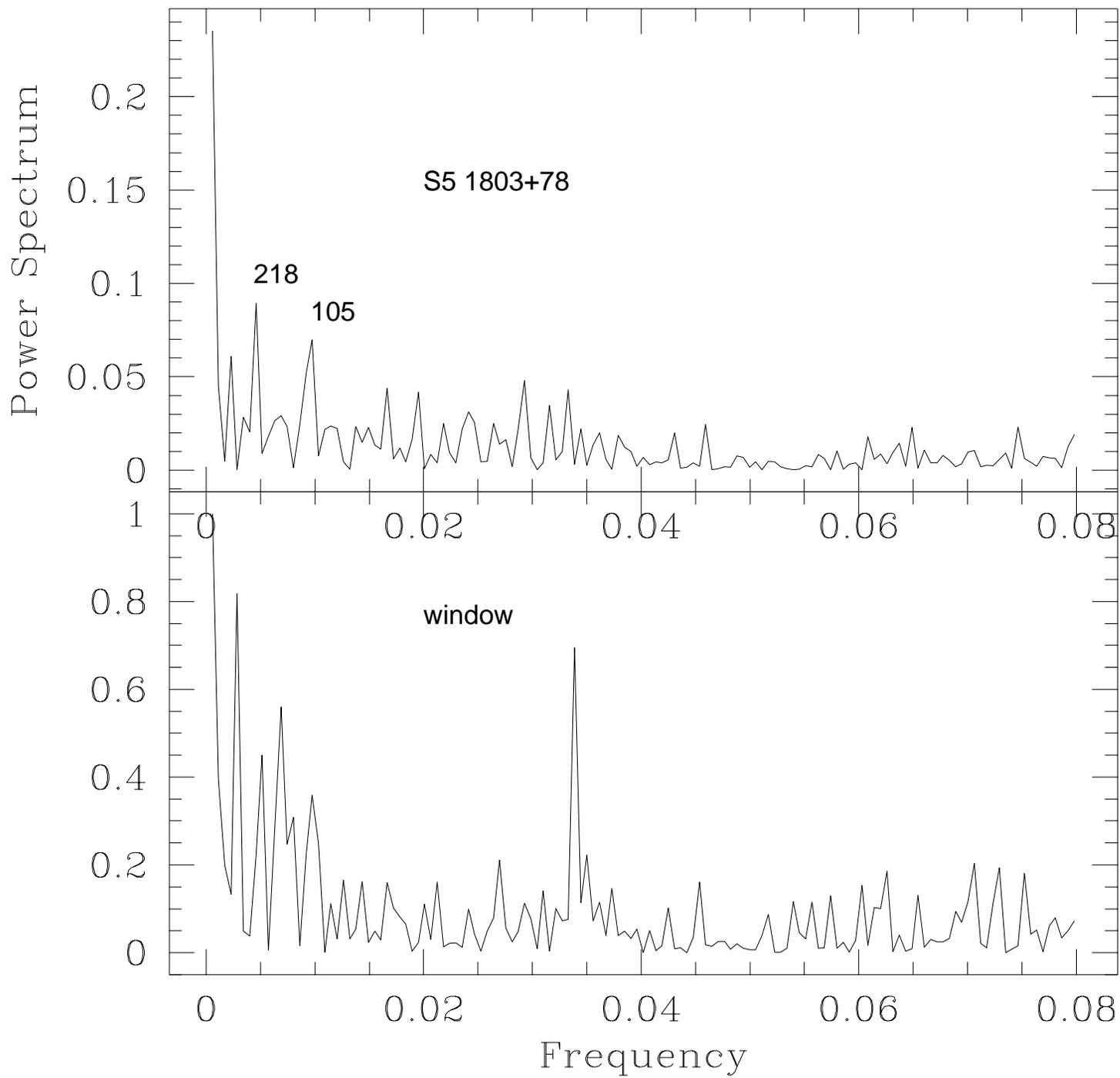
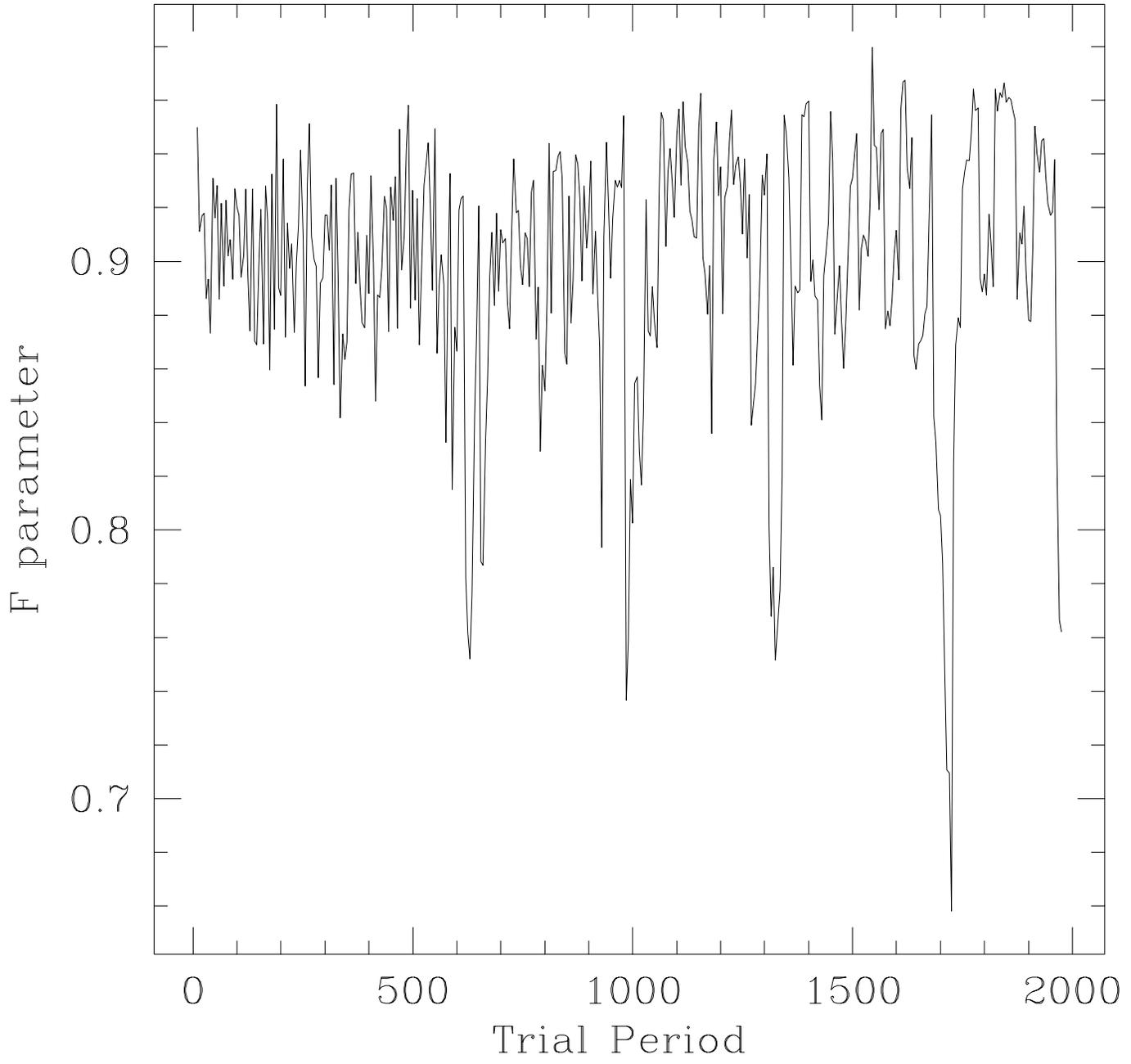
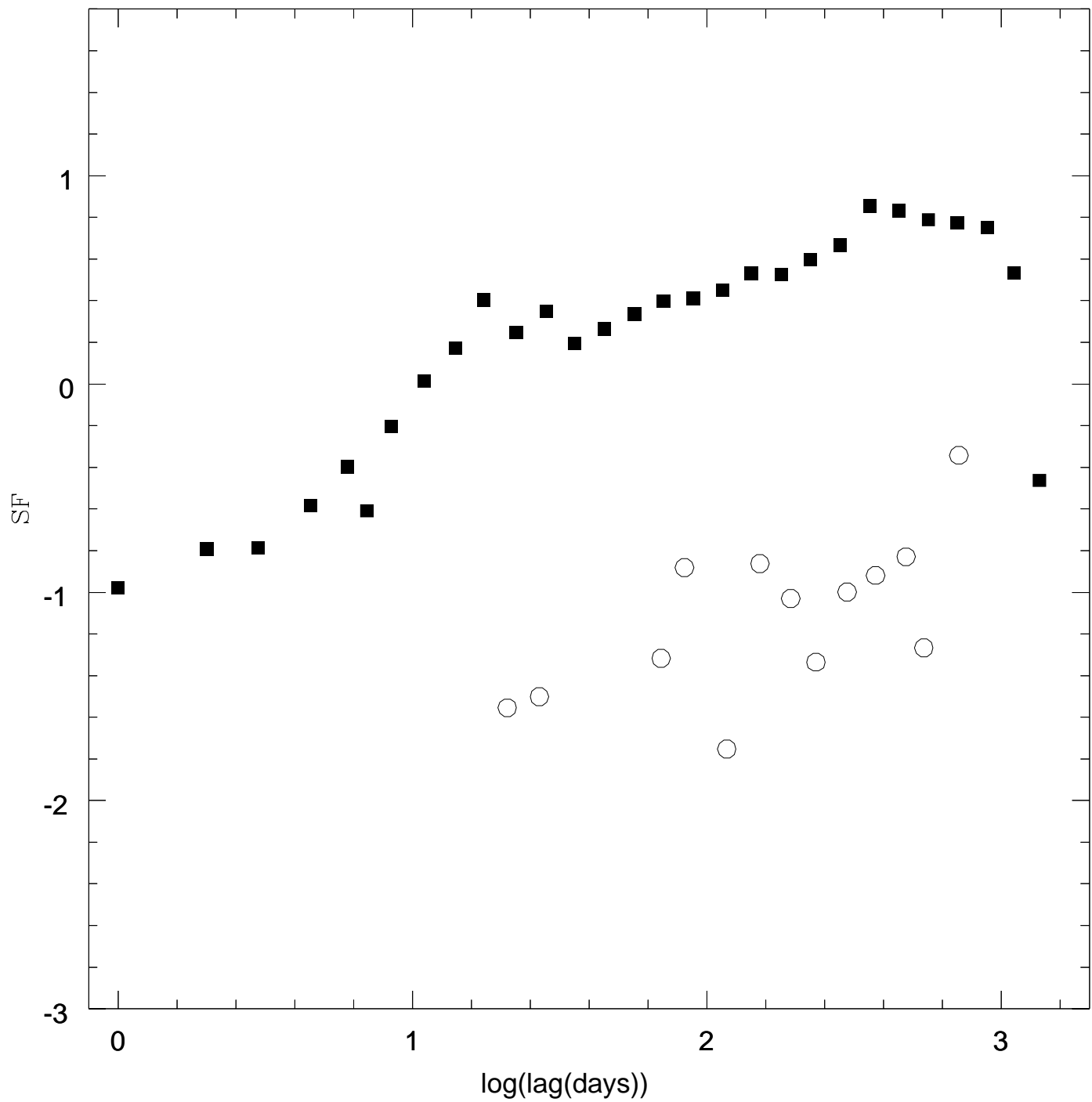
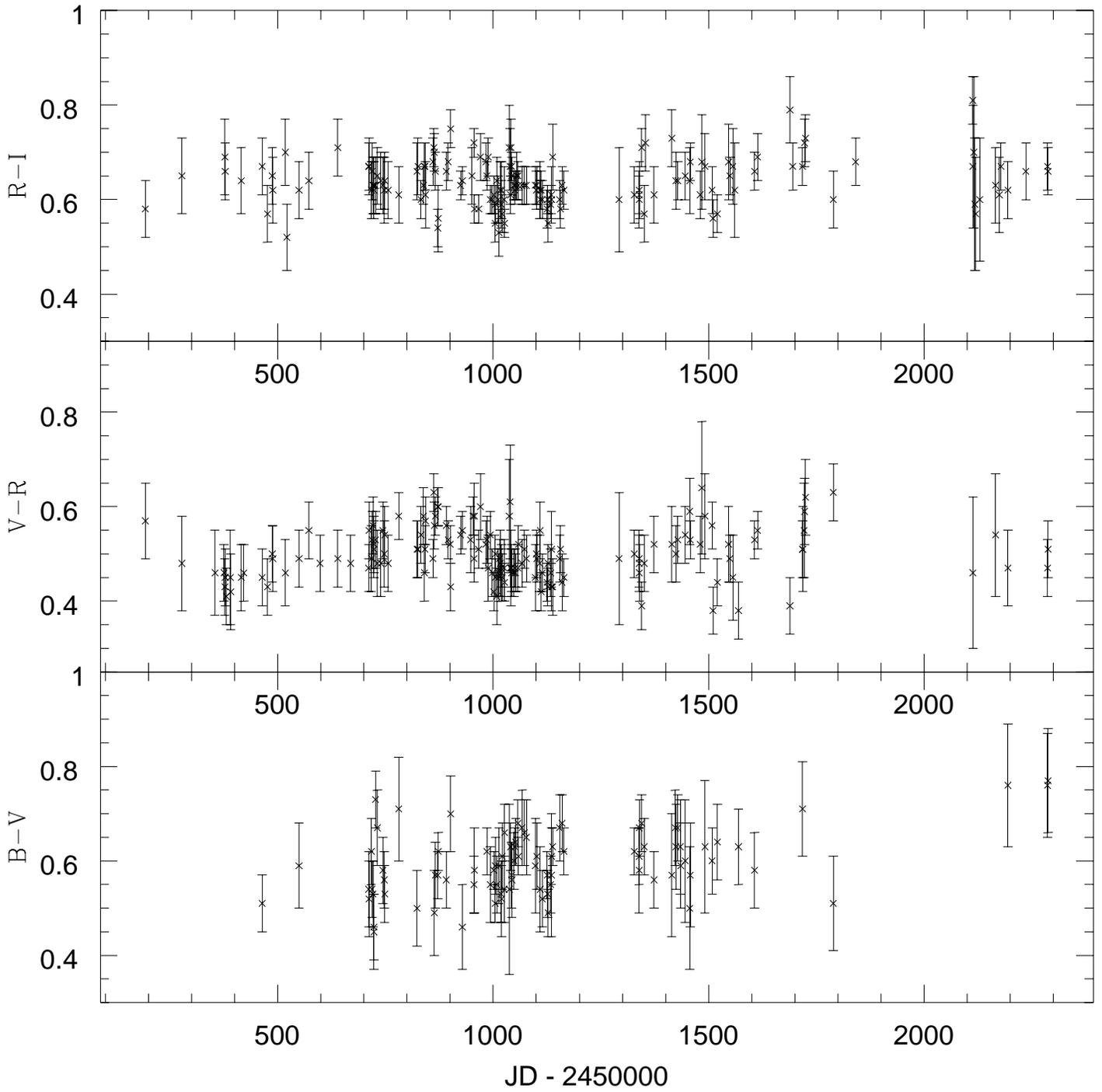
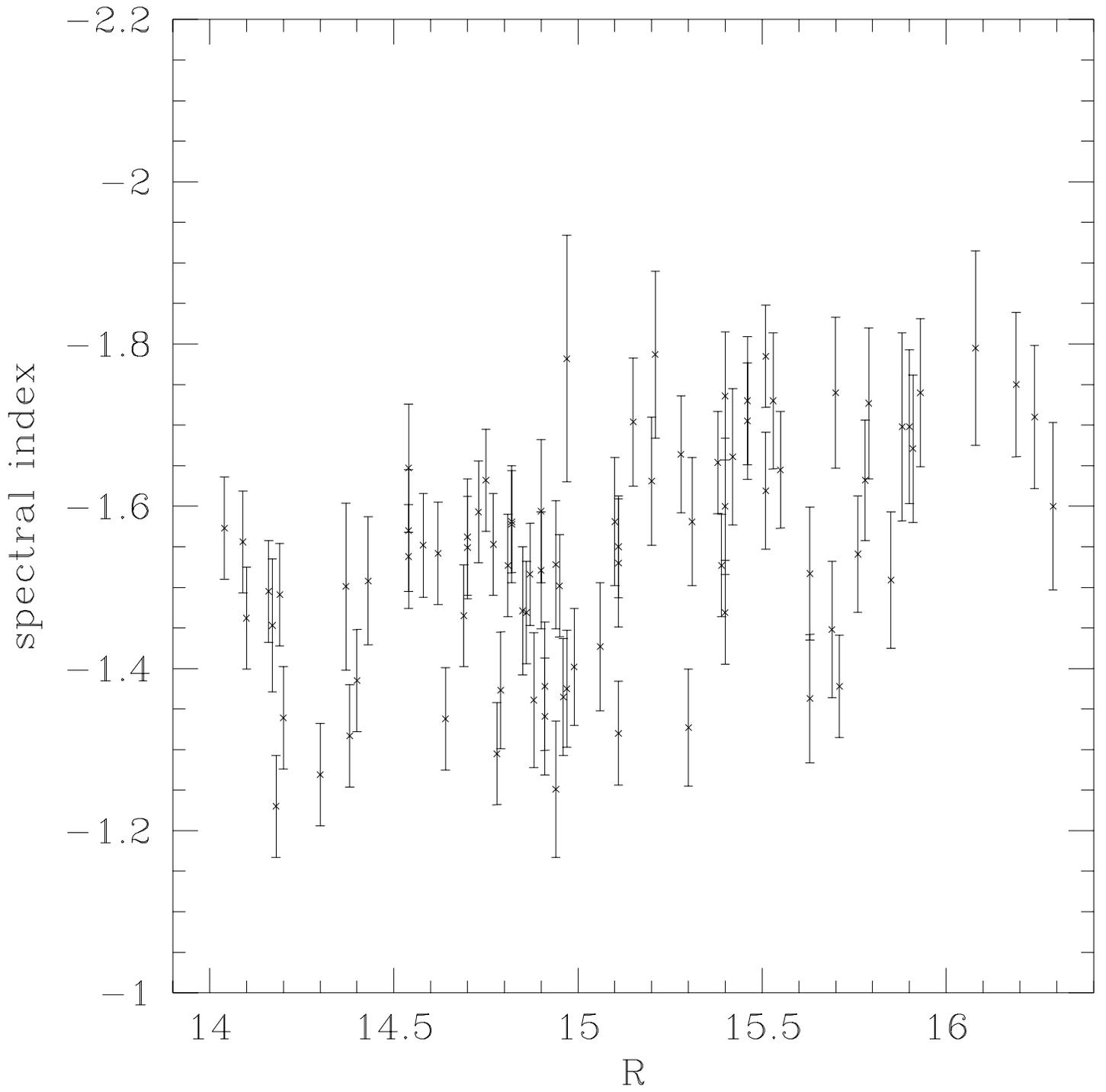
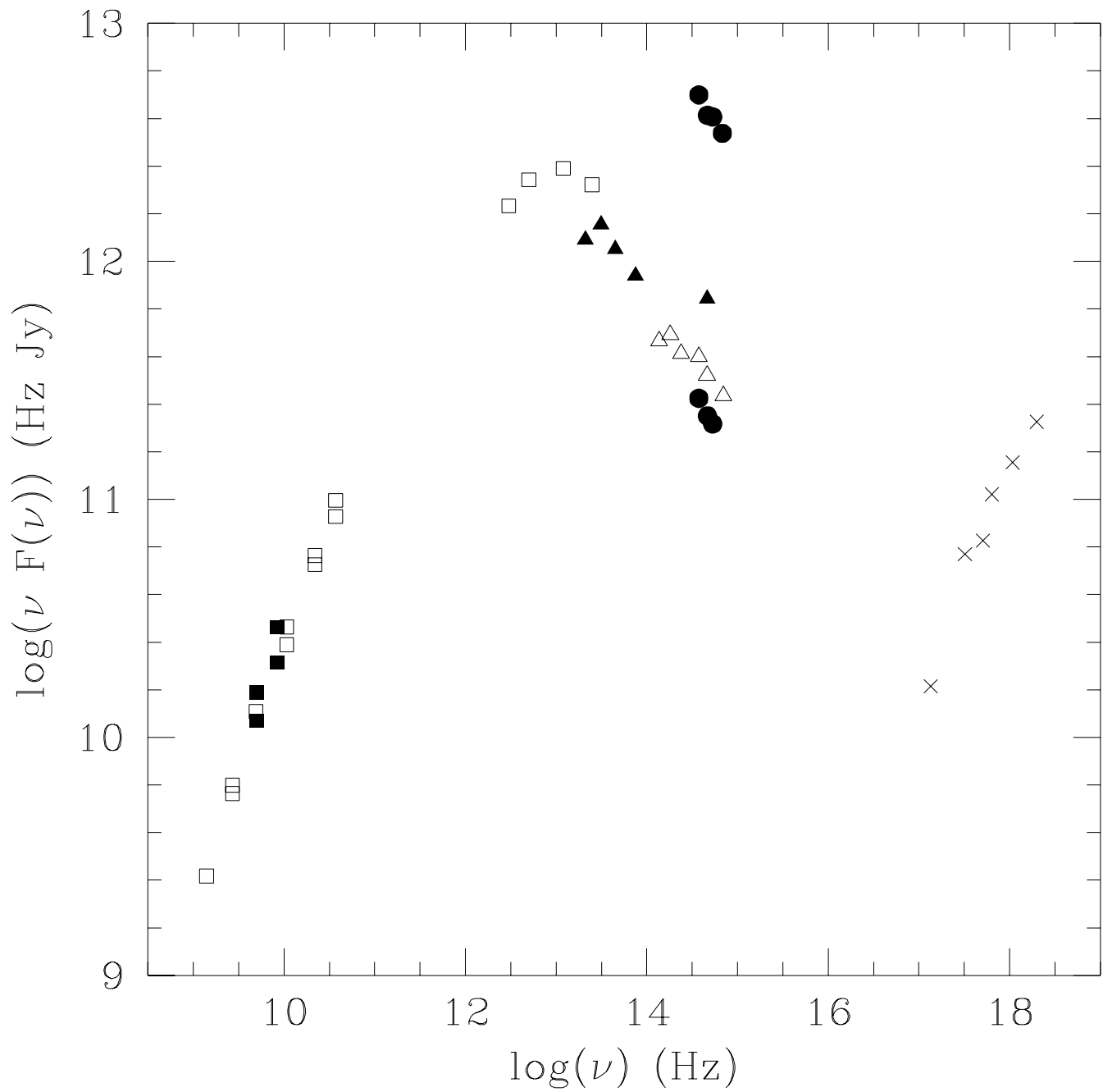
\end{document}